\documentclass[12pt]{article}
\usepackage{a4wide,amssymb,cite}
\pdfoutput=1
\usepackage{graphicx}
\usepackage{a4wide,amssymb,cite}
\usepackage[usenames,dvipsnames]{color}
\usepackage{slashed}
\newcommand{\be}{\begin{equation}}
\newcommand{\ee}{\end{equation}}
\newcommand{\bea}{\begin{eqnarray}}
\newcommand{\eea}{\end{eqnarray}}

\def\circa#1{\,\raise.3ex\hbox{$#1$\kern-.75em\lower1ex\hbox{$\sim$}}\,}

\begin{document}

\begin{titlepage}
%
%


%

\begin{centering}
\vspace{1cm}
{\Large {\bf SIMP dark matter with gauged $Z_3$ symmetry }} \\

\vspace{1.5cm}

{\bf Soo-Min Choi$^*$ and Hyun Min Lee$^\dagger$}
\vspace{.5cm}

{\it Department of Physics, Chung-Ang University, Seoul 156-756, Korea.} 
\\

\end{centering}
\vspace{2cm}

\begin{abstract}
\noindent
We consider a complex scalar field as SIMP dark matter in models with gauged $Z_3$ discrete symmetry appearing as a remnant of dark local $U(1)$.  Dark matter (DM) annihilates dominantly by the 3-to-2 scattering, due to the DM cubic coupling in combination with the DM quartic coupling or the $Z'$ gauge and dark Higgs couplings. We show that a light $Z'$ gauge boson makes DM in kinetic equilibrium with thermal plasma at freeze-out and it affects the DM relic density and perturbativity/unitarity constraints for DM self-interactions. We show that the large DM self-interactions are consistent with solving small-scale structure problems and explaining the DM halo separation recently observed in Abell 3827 cluster. Various bounds on the model from the SIMP conditions, DM self-interactions, $Z'$ searches, DM direct/indirect detection experiments, and Higgs signals, are also discussed.

\end{abstract}

\vspace{4cm}

\begin{flushleft}
$^*$Email: sm90515@cau.ac.kr \\
$^\dagger$Email: hminlee@cau.ac.kr 
\end{flushleft}

\end{titlepage}

\section{Introduction}

It is known that dark matter (DM) makes up about 85\% of the matter density in the Universe \cite{planck} but the identity of dark matter has been regarded as one of the big mysteries in particle physics and cosmology communities. The Weakly Interacting Massive Particles (WIMP) paradigm has driven most of DM searches in terrestrial and satellite experiments, and typically, the abundance of WIMP is determined from the thermal number density at the time of freeze-out in terms of weak-scale DM mass and weak interactions with the SM particles. 
However,  there has been no direct evidence for dark matter other than gravitational or cosmological effects and there are strong limits on the spin-independent DM-nucleon cross section \cite{xenon100,lux}.  Therefore, it is also important to look for alternative testable scenarios for dark matter other than WIMP.

Strong Interacting Massive Particles (SIMP) \cite{simp,simp2} have recently drawn attention due to the interesting property that a sub-GeV dark matter can be thermally produced due to (effective) 5-point self-interactions. SIMP dark matter might provide a new arena for the model building for dark matter that can be testable in the future dark matter searches. Large self-interactions of SIMP dark matter can play a role at the galaxy scales, for instance, for solving the cusp-core and too-big-to-fail problems \cite{small-scale,smallscale2} and explaining the recently observed evidence for self-interacting dark matter in Abell 3827 cluster, namely, the separation between the dark matter halo and the galaxy's stars \cite{abell,raidal,kai}.  Furthermore, the interactions of a light mediator between SIMP and the SM particles can be probed by various indirect and direct experiments \cite{simp2}.

In this article, as a SIMP dark matter candidate, we consider a complex scalar charged under a dark local $U(1)_V$, that is spontaneously broken down to a discrete $Z_3$ group, after another complex scalar gets a nonzero VEV. The remaining $Z_3$ symmetry ensures the stability of dark matter and generates the cubic self-interaction of dark matter dynamically due to the $U(1)_V$ breaking. As  a consequence, dark matter can self-annihilate through the 3-to-2 scattering processes, thanks to the help of DM quartic coupling or $Z'$ gauge and dark Higgs coupling. The dark Higgs could be a potential mediator to keep dark matter in thermal equilibrium via a mixing with the SM Higgs but it turns out to be insufficient due to the smallness of Higgs Yukawa couplings to the SM fermions at the time of SIMP freeze-out.
Instead, the $Z'$ gauge boson or dark photon can mediate for DM to scatter off the SM fermions sufficiently in the presence of a gauge kinetic mixing \cite{simp2}.

We show that the $Z'$ gauge boson also gives sizable contributions to the DM annihilation and scattering processes for most of natural parameters of the model, while helping to realize SIMP as a thermal dark matter. 
We constrain the parameter space of our SIMP dark matter by considering the bounds coming from Bullet cluster \cite{bullet} and halo shapes \cite{haloshape} and find that our model can accommodate a potential hint for self-interacting dark matter from Abell 3827 cluster as estimated in Ref.~\cite{kai}. Various limits from direct and indirect searches for a light $Z'$ gauge boson are considered and compared to the SIMP conditions on the model.   
The direct detection of SIMP dark matter with electron recoil \cite{xenon10,zurek} and the bound from indirect detection and the bound from Higgs signals are also briefly discussed.

The paper is organized as follows. 
We begin to describe the model of a complex scalar dark matter with $Z_3$ gauge symmetry.
Then, we present the results of the 3-to-2 annihilation and self-scattering processes of dark matter and consider the bounds on DM self-interactions.
Next we show the complementarity of the SIMP conditions for the $Z'$ searches at colliders and comment on the direct/indirect detection of dark matter and the impact on Higgs signals. 
Finally, conclusions are drawn. 
There are two appendices summarizing the Higgs-portal and $Z'$-portal interactions and the general formulas for annihilation and scattering cross sections.

\section{The model}

We consider dark matter as a complex scalar $\chi$ having a charge $q_\chi=+1$ under the local $U(1)_V$ symmetry, which is spontaneously broken by the VEV of another complex scalar $\phi$ with charge $q_\phi=+3$.  Thus, the remaining discrete $Z_3$ symmetry is of the gauge symmetry origin \cite{zn,batell} and ensures the stability of scalar dark matter $\chi$.  The SM particles are assumed to be neutral under $U(1)_V$. The $U(1)_V$ charges are summarized in Table 1.

\begin{table}[ht]
\centering
\begin{tabular}{|c||c|c|}
\hline 
& $\phi$ & $\chi$   \\ [0.5ex]
\hline 
$U(1)_V$ & $+3$ & $+1$ 
 \\ [0.5ex]
\hline
\end{tabular}
\caption{$U(1)_V$ charges.}
\label{table:charges1}
\end{table}

The Lagrangian for those scalars and the SM Higgs doublet $H$ is given \footnote{The same local $U(1)_V$ model has been considered for WIMP dark matter in Ref.~\cite{localz3} and the global $U(1)$ case has been also studied in the context of WIMP \cite{globalz31} or SIMP \cite{globalz3} dark matter. } by
\bea
{\cal L}=-\frac{1}{4}V_{\mu\nu} V^{\mu\nu}-\frac{1}{2}\sin\xi \,V_{\mu\nu} B^{\mu\nu}+ |D_\mu\phi|^2 + |D_\mu \chi|^2 + |D_\mu H|^2 - V(\phi,\chi,H)
\eea
where the field strength tensor for dark photon is $V_{\mu\nu}=\partial_\mu V_\nu-\partial_\nu V_\mu$, and covariant derivatives are $D_\mu \phi=(\partial_\mu - iq_\phi g_D V_\mu)\phi$,  
$D_\mu\chi = (\partial_\mu - i q_\chi g_D V_\mu)\chi$, and $D_\mu H= (\partial_\mu - ig' Y_H B_\mu -\frac{1}{2} i g T^a W^a_\mu) H$, and the gauge kinetic mixing between dark photon  $V_\mu$and hypercharge gauge boson $B_\mu$ is introduced by $\sin\xi$.  Then, the dark photon communicates between dark matter and the SM particles through the gauge kinetic mixing. 

The scalar potential is $V(\phi,\chi,H)= V_{\rm DM}+V_{\rm SM}$
with
\bea
V_{\rm DM}&=& -m^2_\phi |\phi|^2 + m^2_{\chi} |\chi|^2 +\lambda_\phi |\phi|^4 +\lambda_\chi |\chi|^4 + \lambda_{\phi \chi} |\phi|^2 |\chi|^2\nonumber \\
&&+\bigg(\frac{\sqrt{2}}{3!}\,\kappa \phi^\dagger \chi^3+{\rm h.c.} \bigg)+ \lambda_{\phi H}|\phi|^2 |H|^2 + \lambda_{\chi H} |\chi|^2 |H|^2, \\
V_{\rm SM}&=&  -m^2_H |H|^2 +\lambda_H |H|^4. 
\eea
Here, we note that $\lambda_\chi, \kappa$ are dimensionless self-couplings of dark matter, $\lambda_{\phi\chi}$ is the coupling between  dark matter and dark Higgs, and $\lambda_{\phi H},\lambda_{\chi H}$ are Higgs-portal couplings. 
As will be shown in a later discussion on SIMP dark matter, the dark photon couplings to the SM fermions become important for keeping dark matter in kinetic equilibrium with the SM.  
Thus, although the Higgs-portal couplings could be interesting for direct/indirect detection, we focus on the $Z'$-portal in the later discussion. 

For a nonzero VEV of $\phi$ with $\langle \phi\rangle=\frac{1}{\sqrt{2}}v'$, the $U(1)_V$ symmetry is broken to a discrete subgroup $Z_3$ and dark photon gets massive and can mix with the SM Z-boson. 
After expanding the dark Higgs as $\phi=\frac{1}{\sqrt{2}}(v'+h')$, the dark Higgs can mix with the SM Higgs by Higgs-portal interaction, $\lambda_{\phi H}$. 
The interaction terms for dark/SM Higgses and dark photon are given in the appendix A .

\section{Dynamics of SIMP dark matter}

 When dark Higgs and dark photon are heavier than dark matter, a pair of dark matter annihilate into neither a single dark matter $+$ a single dark Higgs (semi-annihilation) nor a pair of dark Higgses/dark photons (pair-annihilation). Nonetheless, dark Higgs and $Z'$ contribute to DM annihilation and scattering processes as intermediate states.  Dark Higgs and $Z'$ can decay into a DM pair dominantly if kinematically allowed. For a sizable dark Higgs-DM coupling $\lambda_{\phi\chi}$ and $Z'$ gauge coupling, dark Higgs and $Z'$ would decay before the SIMP freeze-out, so they would not contribute to the observed DM relic density from their decays. 
 
\begin{figure}
  \begin{center}
   \includegraphics[height=0.12\textwidth]{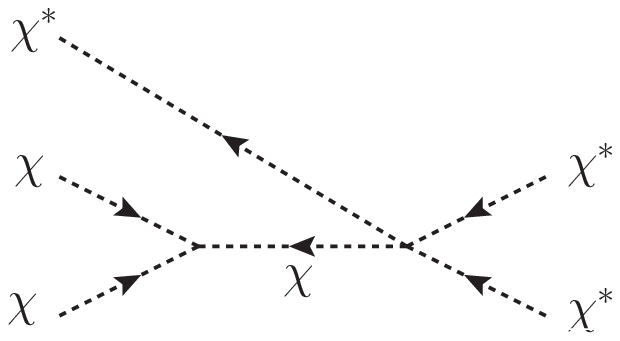}
   \includegraphics[height=0.12\textwidth]{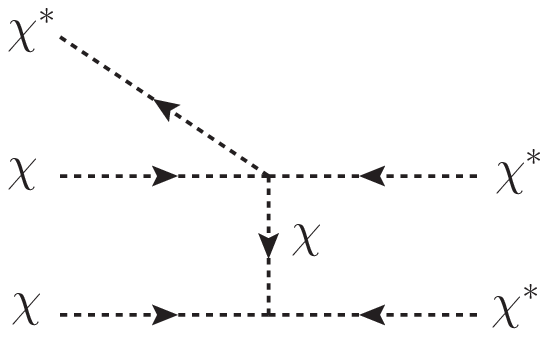}
 \includegraphics[height=0.12\textwidth]{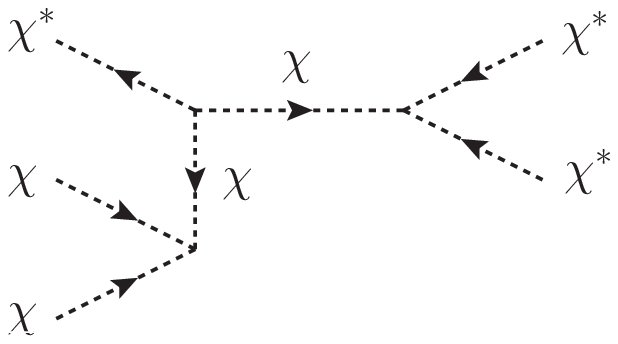}
   \includegraphics[height=0.12\textwidth]{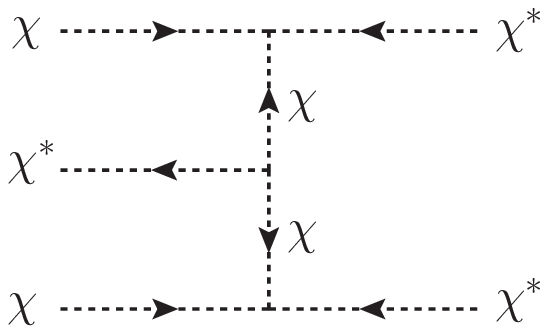}
 \includegraphics[height=0.12\textwidth]{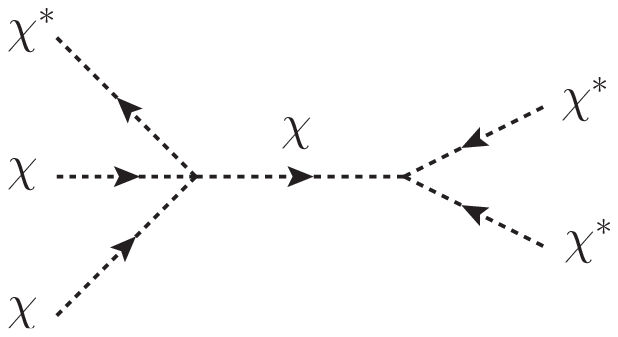}
   \includegraphics[height=0.12\textwidth]{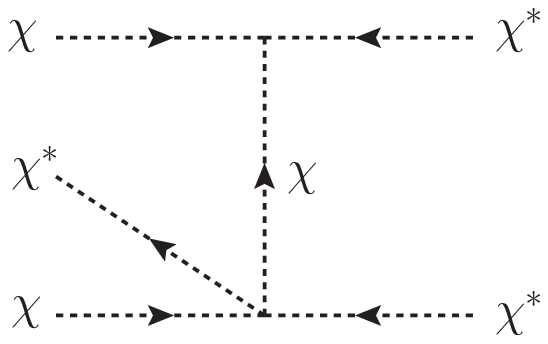}
    \includegraphics[height=0.12\textwidth]{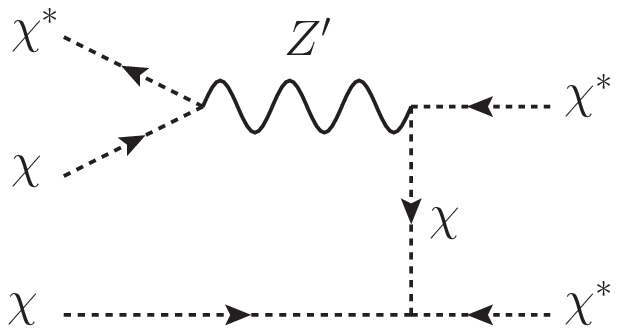}
 \includegraphics[height=0.12\textwidth]{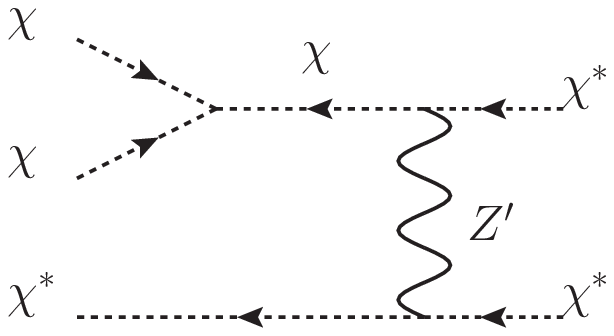}
  \includegraphics[height=0.12\textwidth]{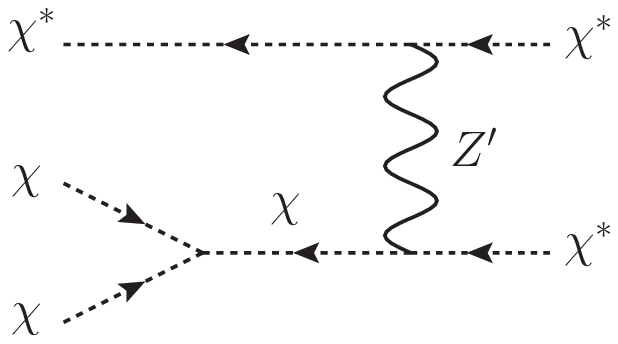}
 \includegraphics[height=0.12\textwidth]{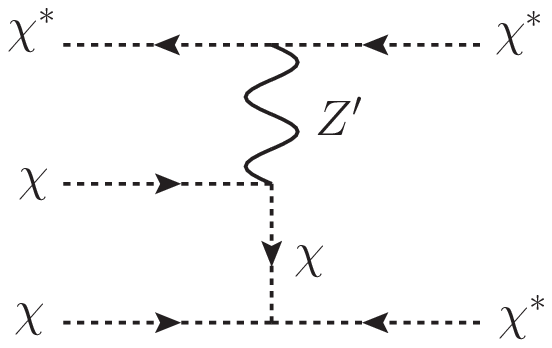}
  \includegraphics[height=0.12\textwidth]{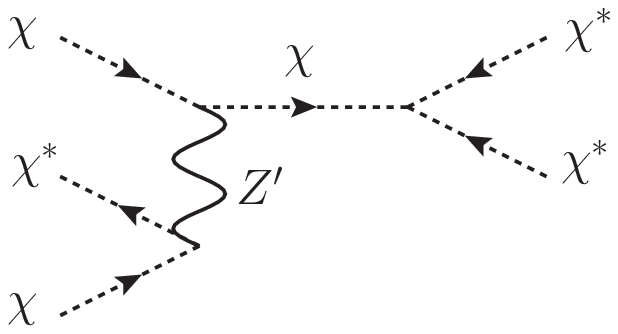}
   \includegraphics[height=0.12\textwidth]{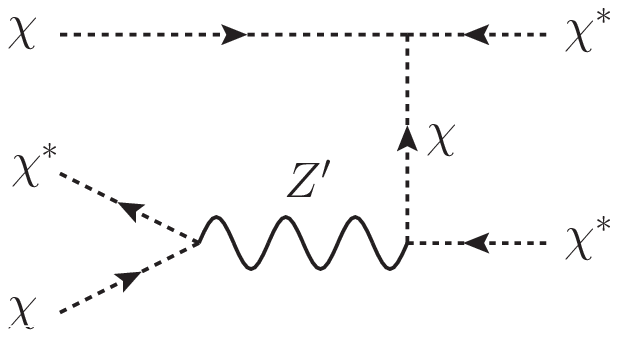}
 \includegraphics[height=0.12\textwidth]{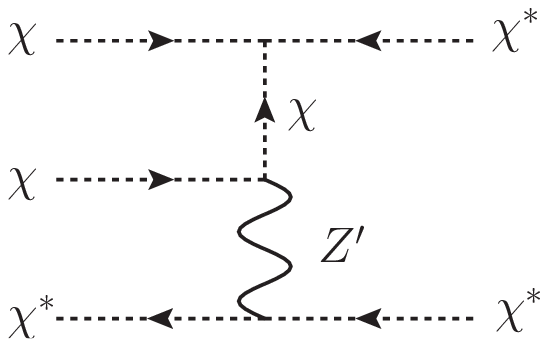}
   \includegraphics[height=0.12\textwidth]{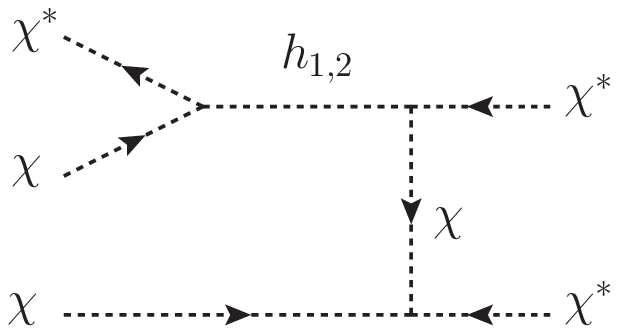}
 \includegraphics[height=0.12\textwidth]{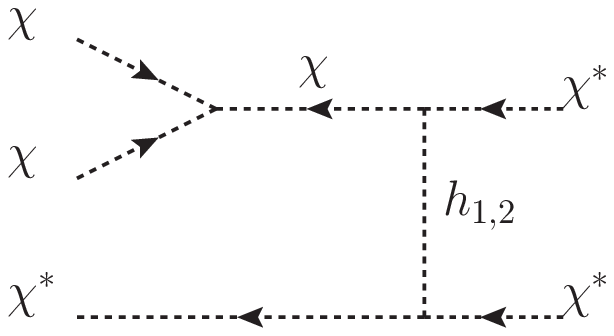}
    \includegraphics[height=0.12\textwidth]{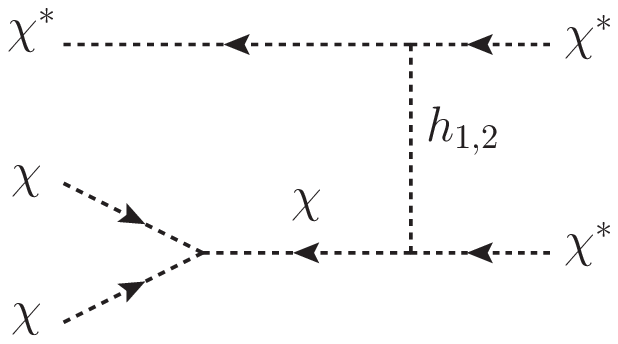}
  \includegraphics[height=0.12\textwidth]{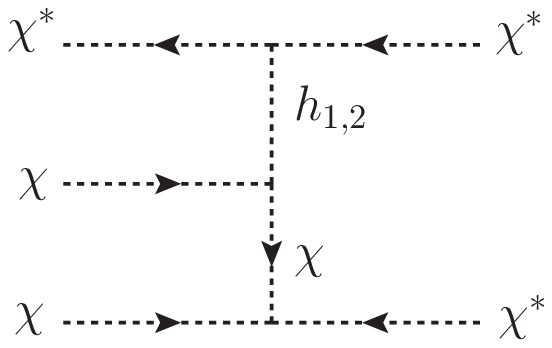}
   \includegraphics[height=0.12\textwidth]{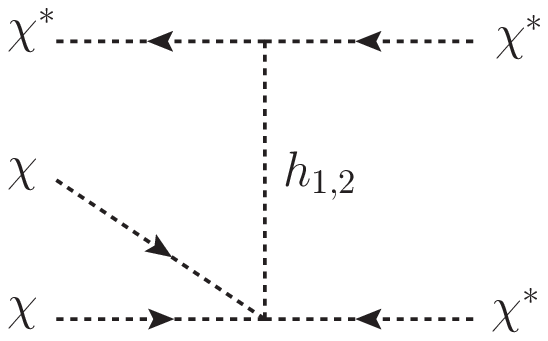}
   \includegraphics[height=0.12\textwidth]{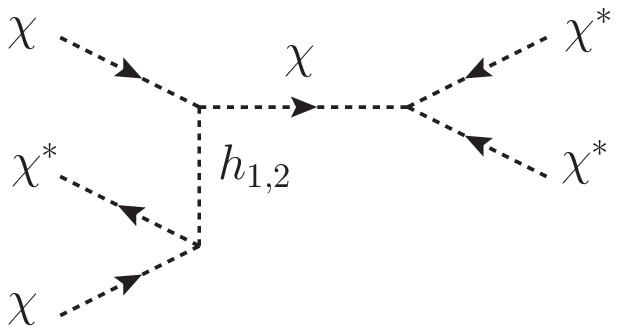}
 \includegraphics[height=0.12\textwidth]{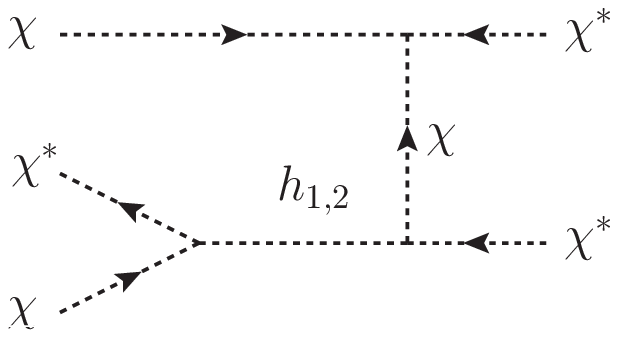}
    \includegraphics[height=0.12\textwidth]{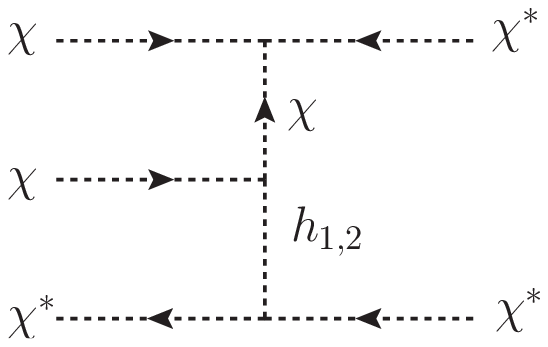}
 \includegraphics[height=0.12\textwidth]{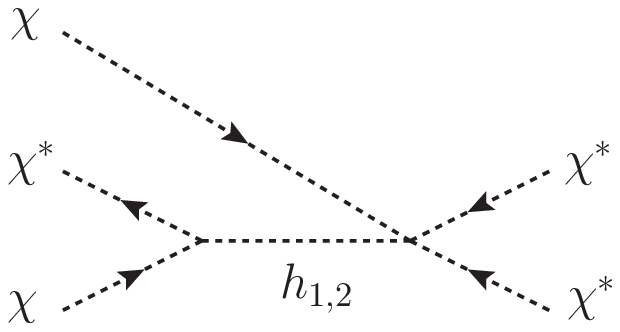}
  \includegraphics[height=0.12\textwidth]{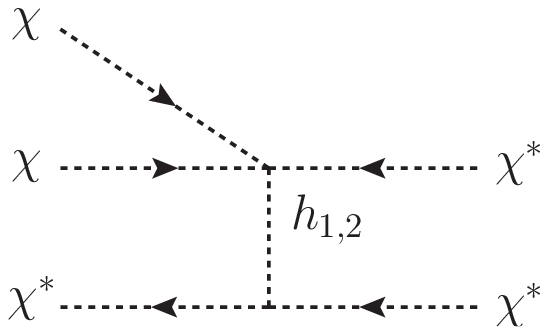}
  \end{center}
  \caption{Feynman diagrams for  $\chi\chi\chi^*\rightarrow \chi^*\chi^*$. }
  \label{3to2a}
\end{figure}

\begin{figure}
  \begin{center}
    \includegraphics[height=0.12\textwidth]{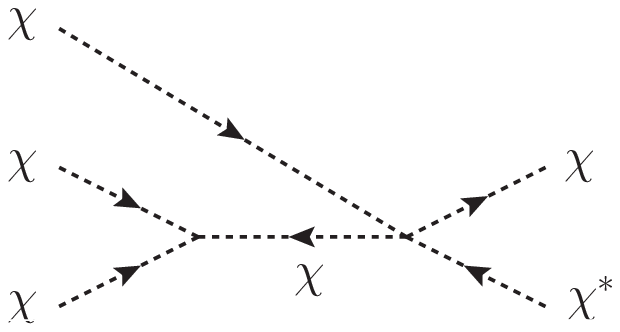}
    \includegraphics[height=0.12\textwidth]{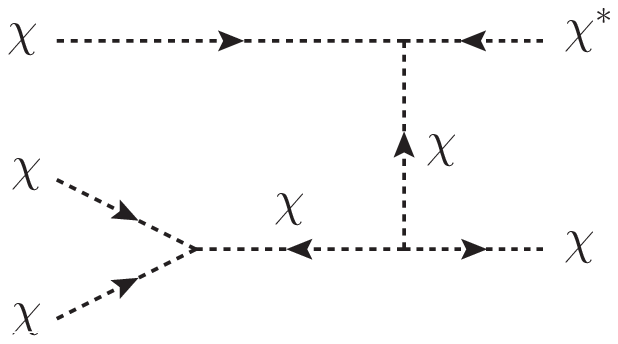}
 \includegraphics[height=0.12\textwidth]{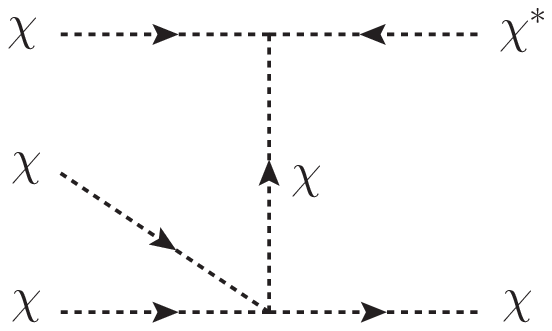}
   \includegraphics[height=0.12\textwidth]{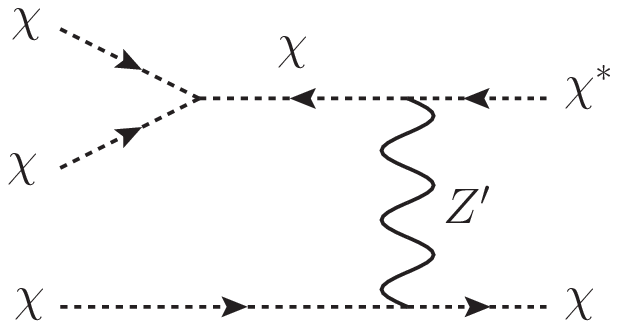}
   \includegraphics[height=0.12\textwidth]{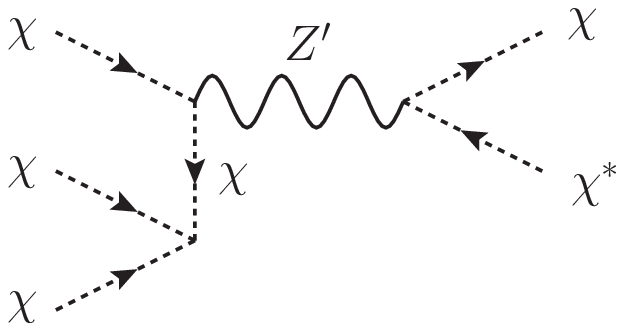}
    \includegraphics[height=0.12\textwidth]{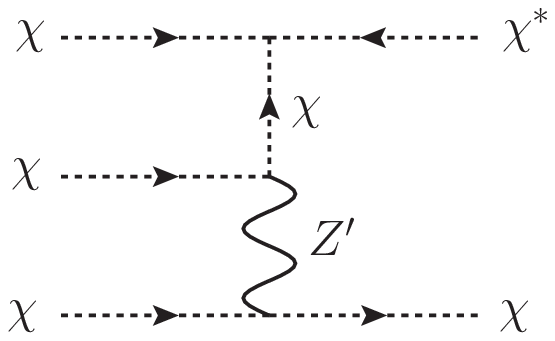}
 \includegraphics[height=0.12\textwidth]{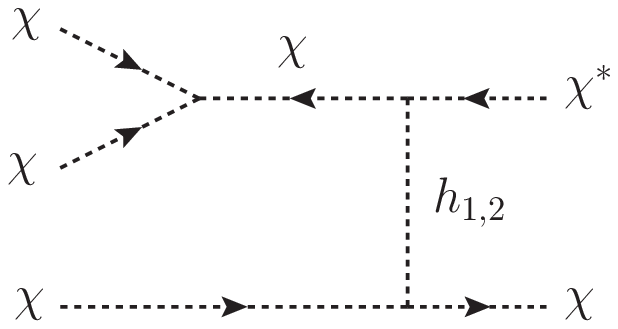}
   \includegraphics[height=0.12\textwidth]{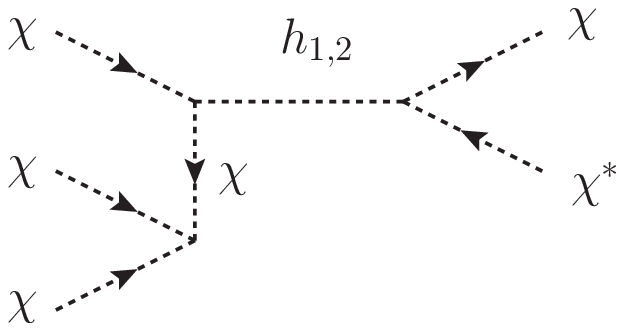}
 \includegraphics[height=0.12\textwidth]{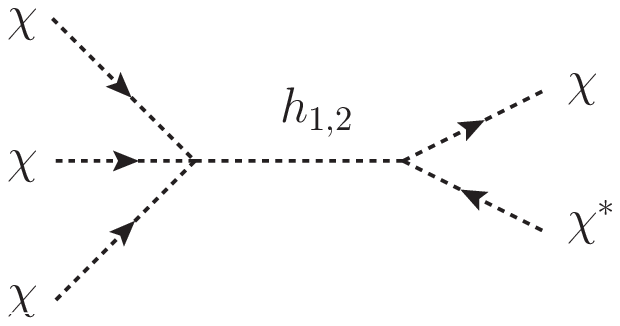}
  \includegraphics[height=0.12\textwidth]{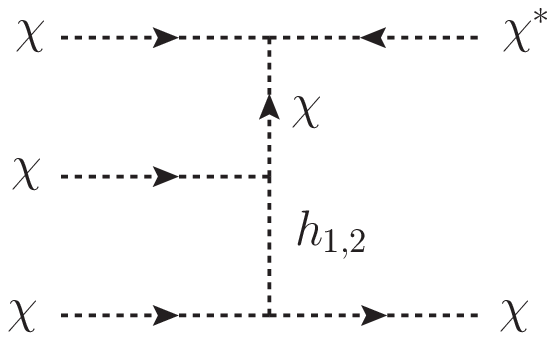}
   \includegraphics[height=0.12\textwidth]{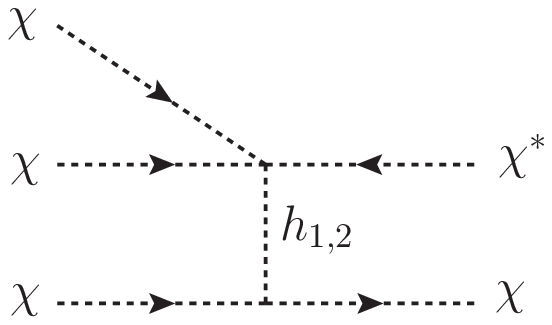}
   \end{center}
  \caption{Feynman diagrams for  $\chi\chi\chi\rightarrow \chi\chi^*$. }
  \label{3to2b}
\end{figure}

Due to the remnant $Z_3$ discrete symmetry, dark matter self-annihilates dominantly through the following 3-to-2 self-annihilation processes, $\chi\chi\chi^*\rightarrow \chi^*\chi^*$, $\chi\chi\chi\rightarrow \chi\chi^*$, and their complex conjugates, as shown in Figs. \ref{3to2a} and \ref{3to2b}.  The scattering processes can be mediated by dark/SM Higgses or dark photon/SM $Z$-boson. In the presence of Higgs and/or $Z'$ portal interactions, the 2-to-2 annihilation such as $\chi\chi^*\rightarrow {\bar f}f$ with $f$ being the SM fermion is possible. The effect of the latter process will be discussed in the next section.  We have adapted the CalcHEP
package \cite{calchep} to our model in order to compute the scattering matrix elements of DM annihilation processes.

As will be discussed in a later section, the $Z'$ portal interaction is important for keeping dark matter in kinetic equilibrium with the SM. We consider both $Z'$  and dark Higgs interactions to DM 
in computing the 3-to-2 annihilation cross sections, but we ignore the $Z'$ and Higgs portal interactions to the SM particles in this section. 
First, the squared amplitude for $\chi\chi\chi^*\rightarrow \chi^*\chi^*$ scattering is, in the non-relativistic limit, given by
\bea
|{\cal M}_{\chi\chi\chi^*\rightarrow \chi^*\chi^*}|^2 &=&
\frac{R^2}{16m^2_\chi}\, \bigg( 74\lambda_\chi-117 R^2 -\frac{200 g^2_D m^2_\chi}{ m^2_\chi+m^2_{Z'} } \nonumber \\
&&+\frac{24 \lambda_{\phi\chi} m_\chi^2 (3 m_\chi^2 - 
         2 m_{h_1}^2)-  \lambda_{\phi\chi}^2 (43 m_\chi^2 - 37 m_{h_1}^2) m_{Z'}^2/(9g^2_D)}{(4 m_\chi^2 - m_{h_1}^2) (m_\chi^2 + m_{h_1}^2)} \bigg)^2
\eea
where $R\equiv \sqrt{2}\kappa v'/(6m_\chi)$.
Likewise, the squared amplitude for $\chi\chi\chi\rightarrow \chi\chi^*$ scattering is, in the non-relativistic limit, given by
\bea
|{\cal M}_{\chi\chi\chi\rightarrow \chi\chi^*}|^2 &=&\frac{3R^2}{m^2_\chi}
\bigg(2\lambda_\chi+9R^2 +\frac{25 g^2_D m^2_\chi}{ m^2_\chi+m^2_{Z'} }   \nonumber \\
&&+\frac{2 \lambda_{\phi\chi} m_\chi^2 (13 m_\chi^2 - 
      2 m_{h_1}^2)-   \lambda_{\phi\chi}^2 (19 m_\chi^2 - m_{h_1}^2)m_{Z'}^2/(9g^2_D)} {(9 m_\chi^2 - m_{h_1}^2) (m_\chi^2 + m_{h_1}^2)} \bigg)^2
\eea
A nonzero cubic DM self-coupling only can make both 3-to-2 annihilation channels possible. Nonzero DM quartic coupling, dark gauge  and dark Higgs couplings, can reduce or enhance the squared amplitudes, but the parameter space of DM self-couplings becomes strongly constrained due to unitarity bounds as will be discussed later.

Assuming CP conservation in the dark sector,  the Boltzmann equation for dark matter number density, $n_{\rm DM}=n_{\chi}+n_{\chi^*}$, with $n_\chi=n_{\chi^*}$,  is
 \bea
\frac{d n_{\rm DM}}{dt} + 3H n_{\rm DM} &=&  -\frac{1}{4} \left(\langle \sigma v^2_{\rm rel}\rangle_{\chi\chi\chi^*\rightarrow \chi^*\chi^*}+ \langle \sigma v^2_{\rm rel}\rangle_{\chi\chi\chi\rightarrow \chi\chi^*} \right)(n^3_{\rm DM} -n^2_{\rm DM} n^{\rm eq}_{\rm DM} )  \nonumber \\
&&-\frac{1}{2} \langle\sigma v_{\rm rel}\rangle_{\chi\chi^*\rightarrow {\bar f} f} (n^2_{\rm DM}- (n^{\rm eq}_{\rm DM})^2).  
 \eea
 As a consequence, using the general formulas for the cross sections in Appendix B, the effective 3-to-2 annihilation cross section appearing in the above Boltzmann equation is obtained as
 \bea
 \langle\sigma v^2_{\rm rel}\rangle_{3\rightarrow 2}&=&\frac{1}{4}  \left(\langle \sigma v^2_{\rm rel}\rangle_{\chi\chi\chi^*\rightarrow \chi^*\chi^*}+ \langle \sigma v^2_{\rm rel}\rangle_{\chi\chi\chi\rightarrow \chi\chi^*} \right) \nonumber \\
 &=& \frac{\sqrt{5}}{1536\pi m^3_\chi} \bigg( |{\cal M}_{\chi\chi\chi^*\rightarrow \chi^*\chi^*}|^2+ |{\cal M}_{\chi\chi\chi\rightarrow \chi\chi^*}|^2  \bigg) \equiv\frac{\alpha^3_{\rm eff}}{m^5_\chi}. \label{3to2}
 \eea
Here, we note that the 2-to-2 annihilation channels in the dark sector are kinematically forbidden while mediators are assumed to couple to the SM particles weakly enough so that we can ignore  the annihilation of dark matter into the SM particles.

Therefore, when the 3-to-2 annihilation is the dominant process at freeze-out, the relic density is determined by $n^2_{\rm DM}\langle\sigma v^2_{\rm rel}\rangle_{3\rightarrow 2}=H$, leading the relic density condition \cite{simp,simp2} as follows, 
\be
m_\chi = \alpha_{\rm eff}\left(\frac{\kappa^2}{x^4_F} \sqrt{\frac{90}{g_*\pi^2}}\,T^2_{\rm eq} M_P\right)^{1/3} 
\ee
where $\alpha_{\rm eff}$ is the effective coupling of dark matter in the 3-to-2 annihilation defined in eq.~(\ref{3to2}) and $T_{\rm eq}$ is the temperature at matter-radiation equality given by $T_{\rm eq}=0.8\,{\rm eV}$. 
Taking $x_F\simeq 20$, and $g_*=10.75$ for the temperature $1\,{\rm MeV}\lesssim T\lesssim 100\,{\rm MeV}$, and $\kappa\equiv\frac{2\pi^2}{45}g_{*s} c=2.55$ for $c\equiv 0.63 g_{*,{\rm eq}}/g_{*s,{\rm eq}}=0.54$,
the above condition becomes
\be
m_\chi = 0.03\, \alpha_{\rm eff}(T^2_{\rm eq} M_P)^{1/3}.   \label{DMrelic}
\ee
For $\alpha_{\rm eff}=1-10$, we get $m_\chi=35-350\,{\rm MeV}$.  In the next section, we search for the parameter space where $\alpha_{\rm eff}$ is of right magnitude for the SIMP dark matter, being compatible with perturbativity and unitarity and the bounds on the DM self-interactions.
From the relic density condition (\ref{DMrelic}) and the DM effective 3-to-2 interaction $\alpha_{\rm eff} $ given in eq.~(\ref{3to2}), we find one of the DM parameters dependent, i.e., the DM quartic coupling is now a dependent parameter by $|\lambda_\chi|\sim \alpha^{3/2}_{\rm eff}/R\propto m_\chi^{3/2}/R$.

\section{Bounds on SIMP dark matter}

In this section, we consider various bounds on SIMP dark matter, starting with bounds on DM self-interactions and $Z'$-portal couplings to the SM fermions.  
The SIMP conditions provide new constraints on the parameter space of dark photon, being complementary to the $Z'$ searches at colliders. Direct detection of SIMP dark matter with electron recoil, indirect detection bounds and the impact on Higgs signals, are also discussed.

\subsection{Self-scattering cross section}

\begin{figure}
  \begin{center}
 \includegraphics[height=0.12\textwidth]{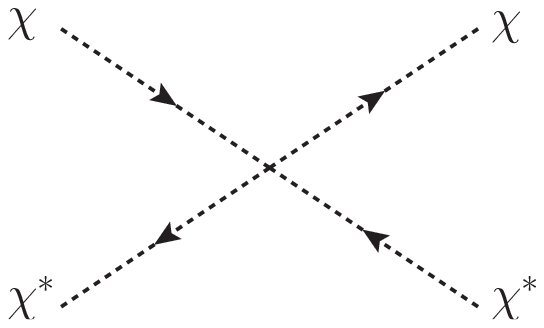}
   \includegraphics[height=0.12\textwidth]{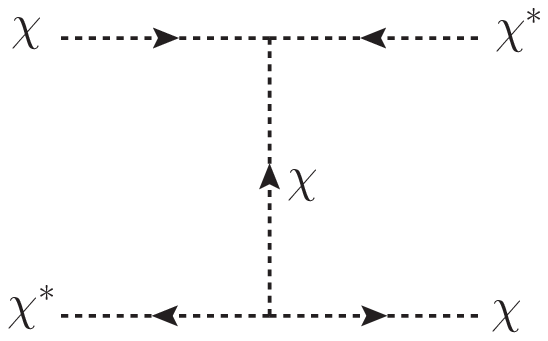}
    \includegraphics[height=0.12\textwidth]{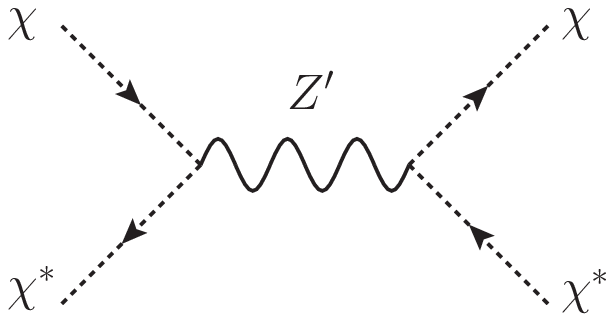}
    \includegraphics[height=0.12\textwidth]{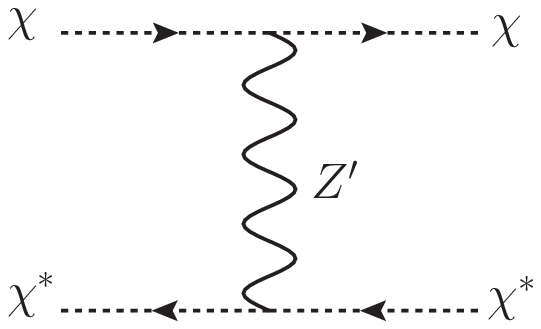}
 \includegraphics[height=0.12\textwidth]{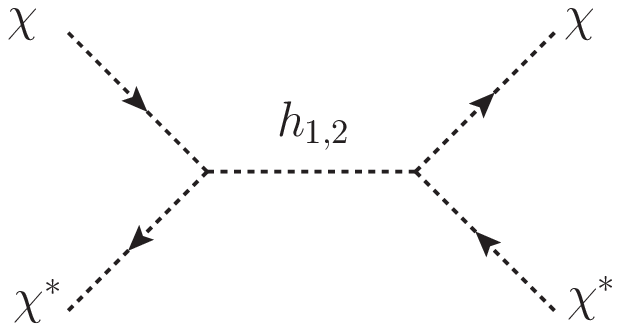}
   \includegraphics[height=0.12\textwidth]{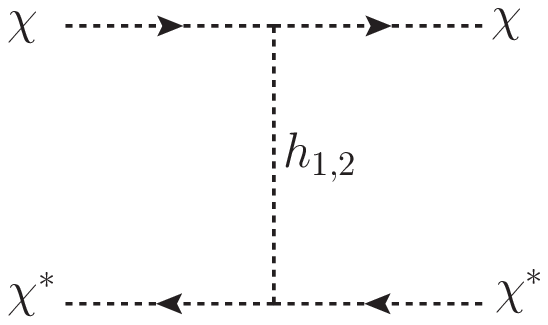}
    \includegraphics[height=0.12\textwidth]{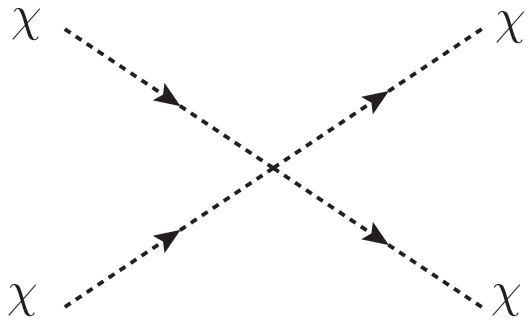}
   \includegraphics[height=0.12\textwidth]{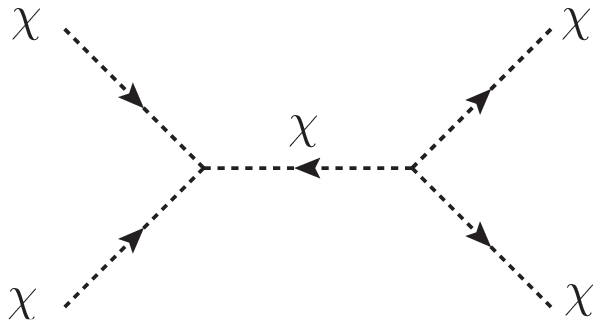}
    \includegraphics[height=0.12\textwidth]{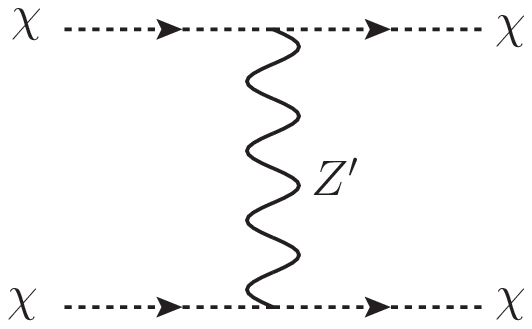}
    \includegraphics[height=0.12\textwidth]{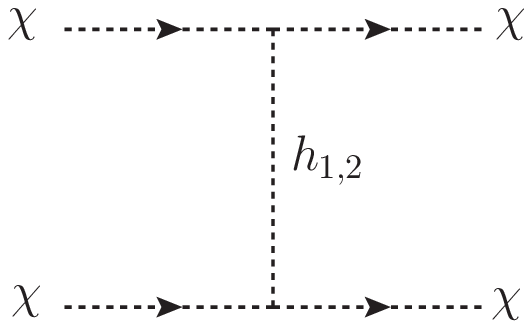}
     \end{center}
  \caption{Feynman diagrams for  $\chi\chi^*\rightarrow \chi\chi^*$ and $\chi\chi\rightarrow \chi\chi$ . }
  \label{2to2}
\end{figure}

Dark matter self-scatters through the following 2-to-2 scattering processes, $\chi\chi\rightarrow \chi\chi$, its complex conjugate, and $\chi\chi^*\rightarrow\chi\chi^*$, as in Fig.~\ref{2to2}. 
Like in the 3-to-2 self-annihilation, we include both dark Higgs and $Z'$ contributions to the scattering processes.

First, the squared amplitude for  the $\chi\chi\rightarrow \chi\chi$ self-scattering is
\be
|{\cal M}_{\rm \chi\chi}|^2=2\left(2\lambda_\chi+ 3R^2+\frac{4g^2_D m^2_\chi}{m^2_{Z'}} -\frac{\lambda_{\phi\chi}^2 m_{Z'}^2}{9g^2_D m^2_{h_1} }  \right)^2.
\ee
On the other hand, the squared amplitude for the $\chi\chi^*\rightarrow \chi\chi^*$ self-scattering is
\bea
|{\cal M}_{\rm \chi\chi^*}|^2=4\left(2\lambda_\chi- 9 R^2-\frac{2g^2_Dm^2_\chi}{m^2_{Z'}}+ \frac{\lambda_{\phi\chi}^2 (2 m_\chi^2 + m_{h_1}^2) m_{Z'}^2}{9g_D^2 (4m_\chi^2 - m_{h_1}^2)m_{h_1}^2} \right)^2.
\eea
Therefore, in the non-relativistic limit for dark matter,  the effective scattering cross section, $\sigma_{\rm self}\equiv \frac{1}{4}(\sigma^{\chi\chi}_{\rm self}+\sigma^{\chi^*\chi^*}_{\rm self}+ \sigma^{\chi\chi^*}_{\rm self} )$ with $\sigma^{\chi^*\chi^*}_{\rm self}=\sigma^{\chi\chi}_{\rm self}$, is
\bea
\sigma_{\rm self} = \frac{1}{64\pi m^2_\chi}\left( |{\cal M}_{\rm \chi\chi}|^2+|{\cal M}_{\rm \chi\chi^*}|^2 \right).
\eea
The self-interaction of dark matter is bounded from $\sigma_{\rm self}/m_{\chi}\lesssim {\rm 1cm^2/g}$ from Bullet cluster \cite{bullet} and halo shapes \cite{haloshape}.  

\begin{figure}
  \begin{center}
    \includegraphics[height=0.45\textwidth]{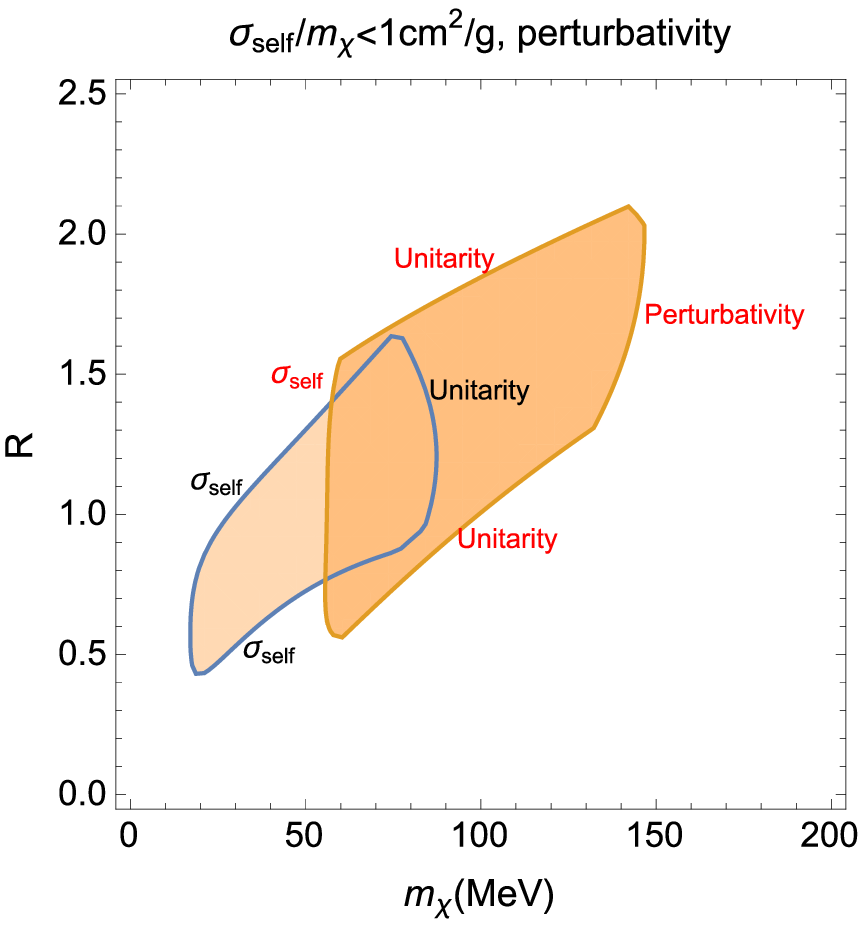}
 \includegraphics[height=0.45\textwidth]{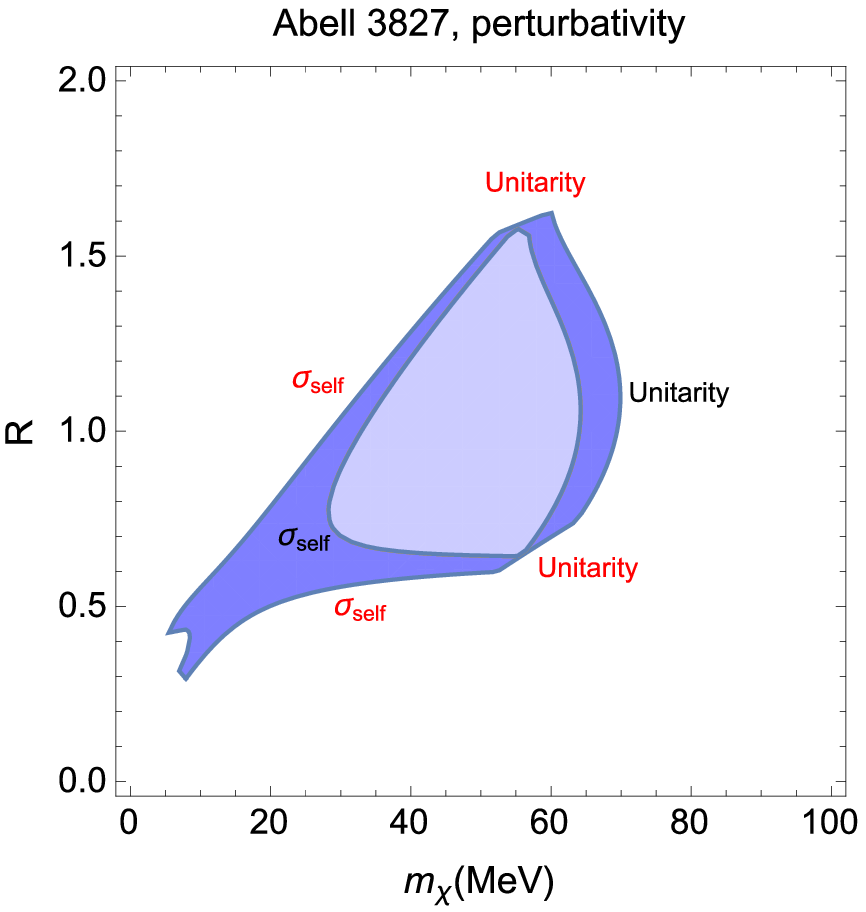}
  \end{center}
  \caption{Left: Parameter space of $m_\chi$ vs $R\equiv \sqrt{2} \kappa v'/(6m_\chi)$, satisfying the relic density, the bounds from Bullet cluster and halo shapes.  $\lambda_{\phi\chi}=0.2, 0.4$ is chosen for light to dark orange regions, respectively. We took $m_{Z'}=5m_\chi$, $g_D=0.1$ and $m_{h_1}=1.5 m_\chi$.
  Right: Parameter space explaining the relic density and the separation
between the dark matter halo and the stars of Abell 3827 with $\sigma_{\rm self}/m_\chi=1-3\,{\rm cm^2/g}$ \cite{kai}. $\lambda_{\phi\chi}=0, 0.1$ is chosen for light to dark blue regions, respectively. In both cases, we took $g_D=0.1$, $m_{Z'}=5m_\chi$ and $m_{h_1}=1.5 m_\chi$, and perturbativity and  unitarity are imposed. In both figures, we assumed a zero Higgs mixing angle, $\sin\theta=0$. Important constraints determining the remaining boundaries of the parameter space are written explicitly in the plots. }
  \label{self}
\end{figure}

We also impose the perturbativity and unitarity bounds on the DM couplings, respectively, as follows,
\bea
\lambda_\chi< 4\pi,  \quad \quad  |{\cal M}_{\chi\chi}|, |{\cal M}_{\chi\chi^*}| < 8\pi. 
\eea
We note that $\chi\chi\rightarrow\chi\chi$ or its complex conjugate determines the unitarity bound dominantly.

\begin{figure}
  \begin{center}
    \includegraphics[height=0.45\textwidth]{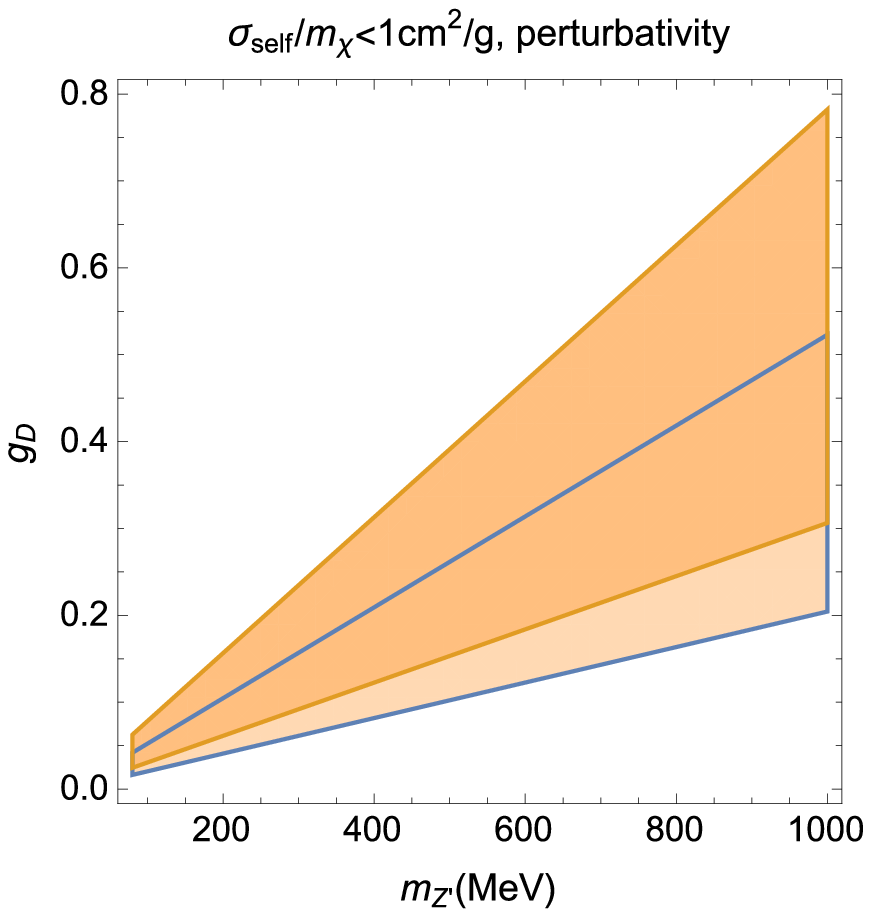}
 \includegraphics[height=0.45\textwidth]{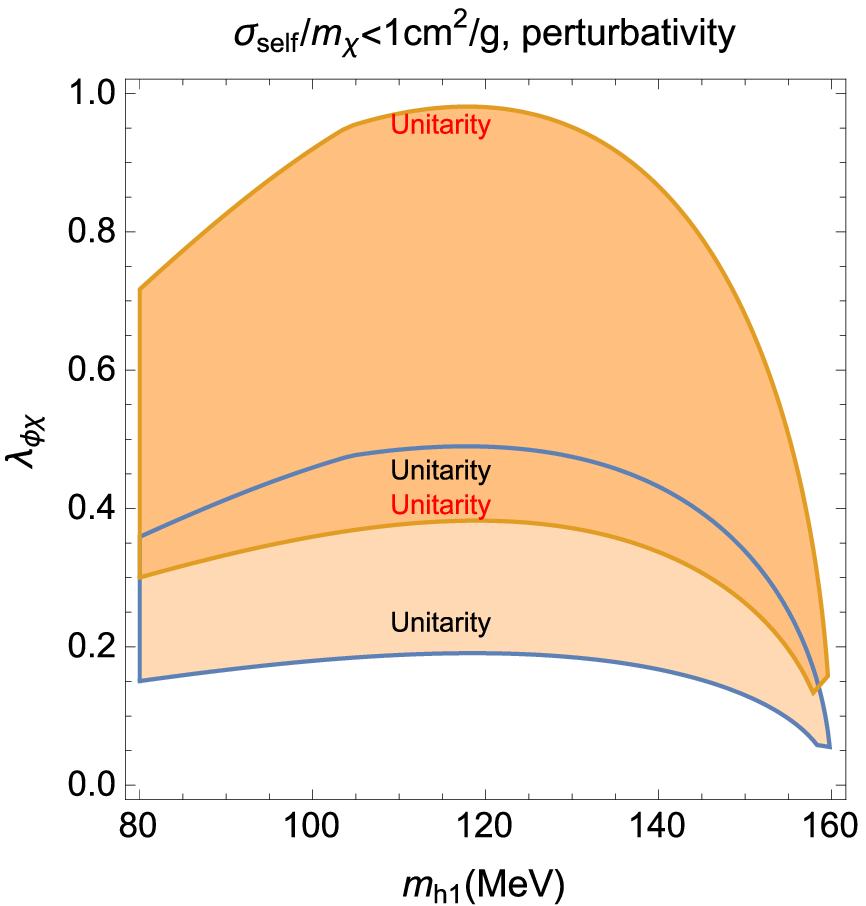}
  \end{center}
  \caption{Left: Parameter space of $m_{Z'}$ vs $g_D$, satisfying the relic density, the bounds from Bullet cluster and halo shapes.  $\lambda_{\phi\chi}=0.4, 0.6$ is chosen for light to dark orange regions, respectively. We took $R=1.5$, $m_\chi=80\,{\rm MeV}$ and $m_{h_1}=120 \,{\rm MeV}$.
  Right: Parameter space of $m_{h_1}$ vs $\lambda_{\phi\chi}$,  satisfying the relic density, the bounds from Bullet cluster and halo shapes.  $g_D=0.1,0.2$ is chosen for light to dark orange regions, respectively. We took $R=1.5$, $m_\chi=80\,{\rm MeV}$ and $m_{Z'}=400\,{\rm MeV}$.  In both figures, we assumed a zero Higgs mixing angle, $\sin\theta=0$. Note that the boundaries of the remaining parameter space are determined by unitarity in both figures. }
  \label{self2}
\end{figure}

\begin{figure}
  \begin{center}
    \includegraphics[height=0.45\textwidth]{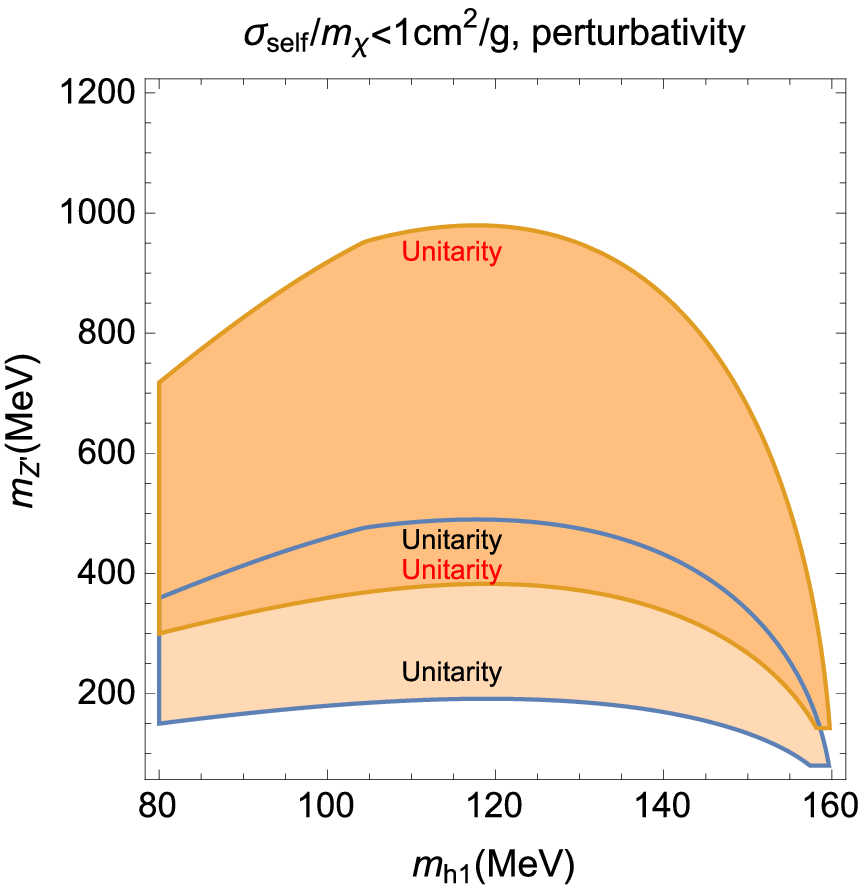}
 \includegraphics[height=0.45\textwidth]{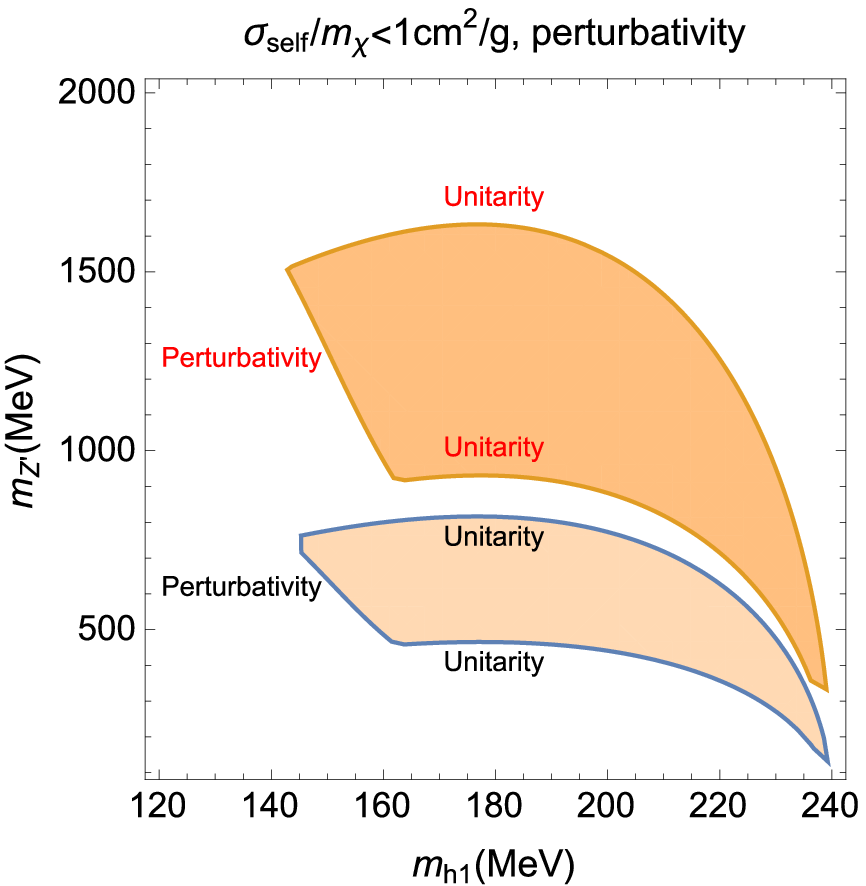}
  \end{center}
  \caption{Parameter space of masses $m_{h_1}$ vs $m_{Z'}$, satisfying the relic density, the bounds from Bullet cluster and halo shapes. We took $m_\chi=80\,{\rm GeV}$ on left and $120\,{\rm MeV}$ on right.  In both figures,  $g_D=0.1, 0.2$ is chosen for light to dark orange regions, respectively, and we took $R=1.5$ and $\lambda_{\phi\chi}=0.4$. We assumed a zero Higgs mixing angle, $\sin\theta=0$.
Important constraints determining the boundaries of the remaining parameter space are written explicitly in the plots.  }
  \label{self3}
\end{figure}

On the left panel of Fig.~\ref{self}, we present the parameter space of DM cubic coupling $\kappa$ and DM mass by considering the relic density, the bounds from Bullet cluster and halo shapes as well as perturbativity and unitarity bounds. Here, we remind that the SIMP relic density condition, eq.~(\ref{DMrelic}), was solved for $\lambda_\chi$, and the self-scattering cross section and perturbativity/unitarity bounds were written in terms of the remaining parameters. The bound on DM self-scattering leads to a lower bound on DM mass on the left of Fig. \ref{self}. On the other hand, the perturbativity and unitarity conditions cut off the parameter space with larger DM masses, because $|\lambda_{\chi}|\sim \alpha^{3/2}_{\rm eff}/R\propto m_\chi^{3/2}/R<4\pi$.  
We find that a relatively light dark Higgs with nonzero $\lambda_{\phi\chi}$  is necessary to satisfy the unitarity bound for $\sigma_{\rm self}/m_{\chi}\lesssim {\rm 1cm^2/g}$, because it cancels the DM self-coupling and dark gauge coupling contributions to the DM self-scattering amplitude. It turns out that the DM masses compatible with bounds on self-interactions are in the range between $20\,{\rm MeV}$ and $150\,{\rm MeV}$ depending on the dark Higgs and $Z'$ couplings. 
On the other hand, the parameter space that explains the DM halo separation observed in Abell 3827 cluster with $\sigma_{\rm self}/m_\chi=1-3\,{\rm cm^2/g}$ at $2\sigma$ \cite{kai} is also shown on the right panel of Fig.~\ref{self}. In this case, the DM self-interaction required for Abell 3827 can be accommodated for $\lambda_{\phi\chi}=0$, namely, without dark Higgs contribution, and the DM masses range between $30\,{\rm MeV}$ and $60\,{\rm MeV}$.

Keeping the relic density, the bounds from Bullet cluster and halo shapes as well as perturbativity and unitarity bounds, we have searched for the parameter space in dark $Z'$ mass vs gauge coupling on left and dark Higgs mass vs $\lambda_{\phi\chi}$ on right, in Fig.~\ref{self2}.  
We find that $Z'$ mass is bounded from above for a given $Z'$ gauge coupling while a relatively light dark Higgs with sizable $\lambda_{\phi\chi}$ is preferred. 
Lastly, in Fig.~\ref{self3}, we also depict the correlation between $Z'$ and dark Higgs masses, depending on $Z'$ gauge coupling and DM mass.  In all Figs.~4-6, we marked the important constraints determining the boundaries of the remaining parameter space in the plots.

\subsection{$Z'$ portal couplings to the SM}

\begin{figure}
  \begin{center}
    \includegraphics[height=0.12\textwidth]{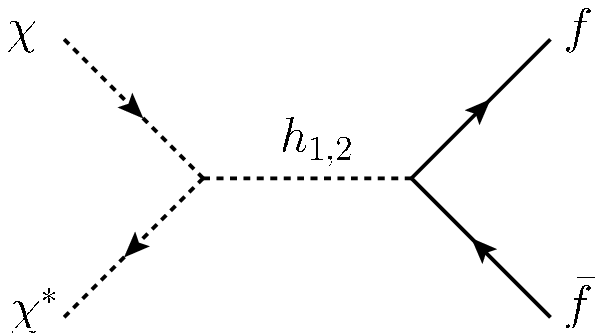}
     \includegraphics[height=0.12\textwidth]{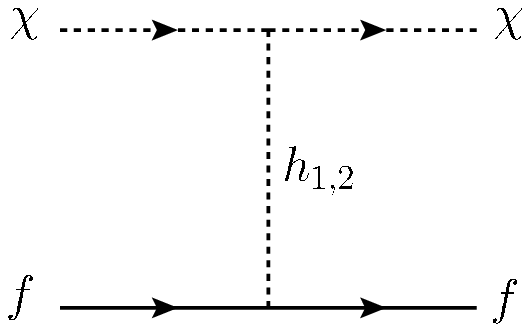} 
 \includegraphics[height=0.12\textwidth]{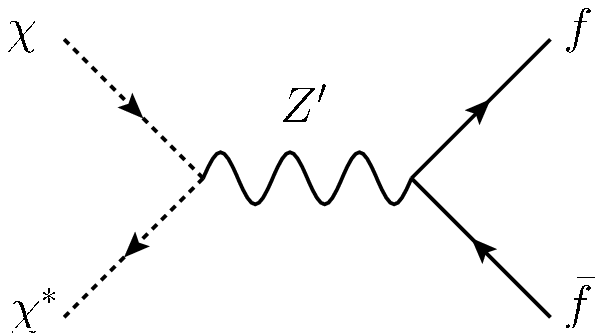}
  \includegraphics[height=0.12\textwidth]{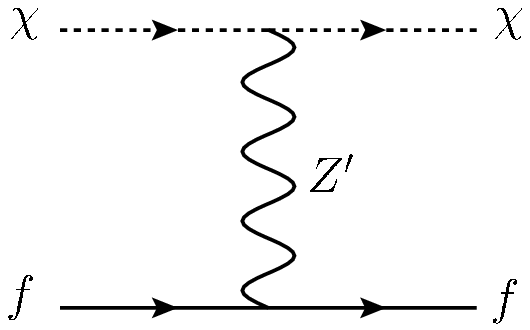}
  \end{center}
  \caption{Dark matter annihilation and scattering through Higgs-portal (Left panel) and $Z'$-portal (Right panel).}
  \label{portals}
\end{figure}

We need a nonzero interaction between dark matter and the SM particles in order for dark matter to keep in kinetic equilibrium until freeze-out, while the consequent DM annihilation into a pair of the SM particles should be subdominant as compared to the 3-to-2 DM annihilation.  These are the SIMP conditions. 
The DM interactions with the SM particles can be induced by Higgs or $Z'$ portals, leading to DM annihilation or scattering as in Fig.~\ref{portals}. In the SIMP scenario, for a sub-GeV DM mass, it was shown that the Higgs-portal coupling is not sufficient for keeping DM in thermal equilibrium \cite{globalz3,simp2}.  
But, for clarification, we present a separate discussion on the Higgs portals in a later section.
In this section, we focus the $Z'$-portal couplings and discuss bounds from SIMP conditions and $Z'$ direct searches.

In the presence of gauge kinetic mixing, dark matter can self-annihilate into a pair of the SM fermions through dark photon exchange in the s-channel as shown in the third Feynman diagram in Fig.~7. The amplitude for $\chi(p_1) \chi^*(p_2)\rightarrow f(k_1) {\bar f}(k_2)$ process is
\bea
i{\cal M}_{\rm ann}=-i g_D (p_1-p_2)^\mu \,\cdot \frac{-i g_{\mu\nu}}{s-m^2_{Z'}+im_{Z'}\Gamma_{Z'}}\,\cdot (-i)e Q_f \varepsilon\, {\bar u}(k_1)\gamma^\nu v(k_2)
\eea
where $Q_f$ is the electromagnetic charge of the SM fermion and $\varepsilon\simeq \cos\theta_W \xi$ for $m_{Z'}\ll m_Z$.  
Then, the squared amplitude is given by
\be
|{\cal M}_{\rm ann}|^2=\frac{4{\varepsilon}^2 Q^2_f e^2 g^2_D}{(s-m^2_\chi)^2} \bigg[ 2(k_1\cdot (p_1-p_2))(k_2\cdot (p_1-p_2))-(k_1\cdot k_2)(p_1-p_2)^2-m^2_f (p_1-p_2)^2 \bigg].
\ee
Therefore, in the non-relativistic limit for dark matter,  the annihilation cross section for $\chi\chi^*\rightarrow f {\bar f}$ with $Q_f=-1$ for charged leptons is
\bea
\langle \sigma v_{\rm rel}\rangle_{\rm ann}&=& \frac{1}{64\pi p^0_1 p^0_2}\int d\Omega\, \bigg\langle \frac{{|{\vec k}_2|}}{\sqrt{s}}|{\cal M}_{\rm ann}|^2\bigg\rangle   \nonumber \\
&=& \frac{2\varepsilon^2 e^2 g^2_D m^2_\chi}{\pi[(4 m^2_\chi-m^2_{Z'})^2+m^2_{Z'} \Gamma^2_{Z'}] }  \bigg(\frac{T}{m_\chi}\bigg)\equiv  \frac{\delta^2_1}{m^2_\chi}.
\label{annsection}
\eea

On the other hand, dark matter can scatter with the SM fermions that are in thermal equilibrium at dark matter freeze-out.  The amplitude for $\chi(p_1) f (p_2)\rightarrow \chi (k_1) f(k_2)$  process as shown in the last Feynman diagram in Fig.~7 is
\bea
i{\cal M}_{\rm scatt}=-i g_D (p_1+k_1)^\mu \,\cdot \frac{-i g_{\mu\nu}}{t-m^2_{Z'}}\,\cdot (-i)eQ_f \varepsilon\, {\bar u}(k_2)\gamma^\nu u(p_2)
\eea
with $t\equiv (p_1-k_1)^2$. 
Then, the spin-averaged amplitude squared is given by
\be
|{\cal M}_{\rm scatt}|^2= \frac{2{\varepsilon}^2 Q^2_f e^2 g^2_D}{(t-m^2_{Z'})^2}\,\bigg[2(k_2\cdot (p_1+k_1))(p_2\cdot (p_1+k_1))-(k_2\cdot p_2)(p_1+k_1)^2+m^2_f (p_1+k_1)^2 \bigg].  \label{melements}
\ee
Around the freeze-out, the SM fermions in thermal equilibrium are relativistic so we can take the momenta relevant for the DM-SM fermion scattering  to be
$p_1=(m_\chi,p,0), p_2=(p,-p,0)$ and $k_1=(m_\chi,p\cos\theta,p\sin\theta)$, $k_2=(p,-p\cos\theta,-p\sin\theta)$ in the center of mass frame where $\theta$ is the scattering angle of dark matter. 
Therefore, we obtain the kinetic scattering cross section for $\chi f \rightarrow \chi f $ with $Q_f=-1$ as
\bea
\langle\sigma v_{\rm rel}\rangle_{\rm scatt}&=&\frac{1}{64\pi p^0_1 p^0_2}\int d\Omega\, \bigg\langle \frac{{|{\vec k}_1|}}{\sqrt{s}}|{\cal M}_{\rm scatt}|^2\bigg\rangle   \nonumber \\
 &=&\frac{3\varepsilon^2 e^2 g^2_D m^2_\chi}{2\pi m^4_{Z'}} \bigg(\frac{T}{m_\chi}\bigg) \equiv  \frac{\delta^2_2}{m^2_\chi}\label{kinsection}.
\eea
The cross section for $\chi {\bar f} \rightarrow \chi {\bar f}$ process is given by the same formula in eq.~(\ref{kinsection}).

\begin{figure}
  \begin{center}
    \includegraphics[height=0.38\textwidth]{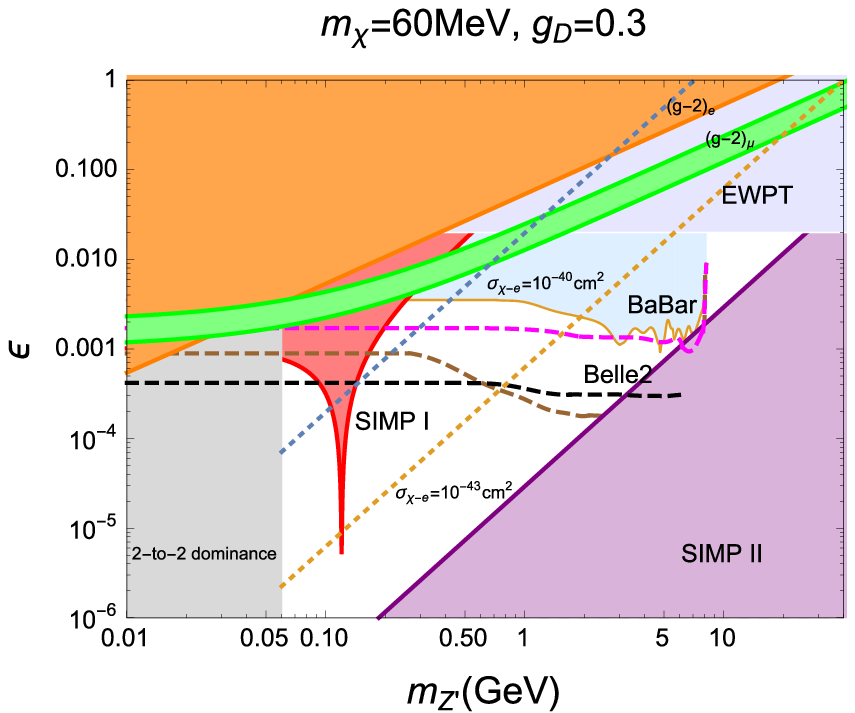}
      \includegraphics[height=0.38\textwidth]{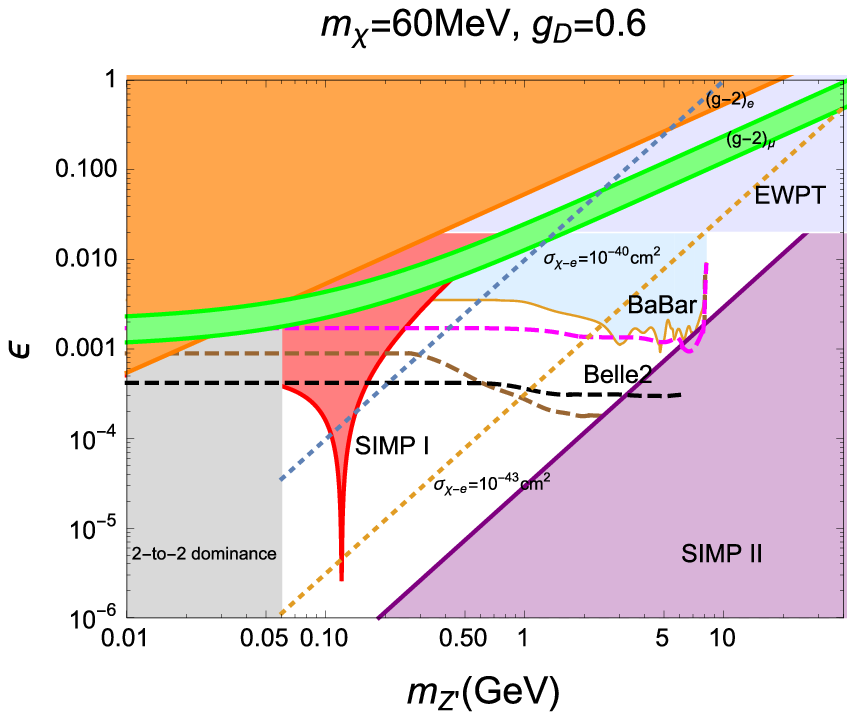}
   \end{center}
  \caption{ Various limits on the parameter space of gauge kinetic mixing $\varepsilon\equiv \xi\cos\theta_W$ vs $m_{Z'}$.  We have chosen $m_\chi=60\,{\rm MeV}$ and $g_D=0.3(0.6)$ on left(right). SIMP I corresponds to  the region that does not fulfill  the condition, $n_{\rm DM} \langle\sigma v_{\rm rel}\rangle_{\rm ann} < n^2_{\rm DM} \langle \sigma v^2_{\rm rel}\rangle_{\rm 3\rightarrow 2}$, while SIMP II corresponds to the region that does not satisfy the  thermal equilibrium condition, $n^2_{\rm DM} \langle \sigma v^2_{\rm rel}\rangle_{\rm 3\rightarrow 2} < n_{\rm SM} \langle \sigma v_{\rm rel}\rangle_{\rm kin}$.}
  \label{zprime}
\end{figure}

The SIMP conditions are imposed at freeze-out temperature, as follows,
\bea
n_{\rm DM} \langle\sigma v_{\rm rel}\rangle_{\rm ann} < n^2_{\rm DM} \langle \sigma v^2_{\rm rel}\rangle_{\rm 3\rightarrow 2} < n_{\rm SM} \langle \sigma v_{\rm rel}\rangle_{\rm kin}.  \label{simp}
\eea
The above conditions (\ref{simp}) constrain the 2-to-2 annihilation cross section and the dark matter kinetic scattering in terms of $\delta_{1,2}$ defined in eqs.~(\ref{annsection}) and (\ref{kinsection}) and $\alpha_{\rm eff}$, as follows, 
\bea
\delta_1 \lesssim 2.4\times 10^{-6}\,\alpha_{\rm eff},     \quad \quad  \delta_2\gtrsim  10^{-9}\alpha^{1/2}_{\rm eff}.
\eea
Then, using eqs.~(\ref{annsection}) and (\ref{kinsection}) with $\frac{T}{m_\chi}\simeq \frac{1}{20}$, the SIMP conditions bound the dark gauge coupling and the gauge kinetic mixing as a function of dark photon mass as
\bea
2.2\times 10^{-8} \alpha^{1/2}_{\rm eff}\,\frac{m^2_{Z'}}{m^2_\chi}  \lesssim |\varepsilon| g_D\lesssim 4.4 \times 10^{-5} \alpha_{\rm eff}\,\sqrt{\bigg(4-\frac{m^2_{Z'}}{m^2_\chi}\bigg)^2+\Gamma^2_{Z'}\frac{m^2_{Z'}}{m^2_\chi} }.
\eea

The parameter space of dark photon  can be constrained by direct searches at colliders \cite{beam dump,babarmet,toro0,prospect,batell2, babarlepton,toro} and lepton $g-2$  experiments \cite{g-2}, etc. There has been a recent study on the limits on light dark matter in the context of a vector messenger \cite{toro}. 
In Fig.~\ref{zprime}, various limits are imposed in the parameter space of the gauge kinetic mixing $\varepsilon\simeq \xi\cos\theta_W$ vs $m_{Z'}$, and some values of DM-electron scattering cross sections are presented.  The direct limits from monophoton $+$ MET at BaBar in light blue\cite{babarmet,toro0,prospect,batell2} and the future prospects at Belle2 in dashed lines \cite{prospect} are given; the indirect limits from electron $g-2$ in orange and EWPT are presented; the muon $g-2$ favored region in green is shown too. 
The SIMP conditions, (SIMP I in red is the dominance of the 2-to-2 annihilation into the SM fermions while SIMP II in purple  is the decoupling of SIMP from thermal plasma), rule out the parameter space that is not covered by direct and indirect limits on dark photon.  
The parameter space with $m_{Z'}<m_\chi$ in gray is ruled out due to the dominance of the 2-to-2 annihilation into dark photons. 

The recent BaBar search for monophoton $+$ dilepton (not shown) \cite{babarlepton} can limit $\varepsilon$ at the level of $10^{-4} - 10^{-3}$ for dark photon masses in the range $0.02 - 10.2$ GeV, but the limits depend on the decay branching fraction of dark photon. When the decay of dark photon into a pair of dark matter is open, it dominates easily, so the previously mentioned monophoton $+$ MET is more constraining. For instance, the monophoton $+$ MET at BaBar applies for dark photon mass being above $120\,{\rm MeV}$ in Fig.~\ref{zprime}, while the monophoton $+$ dilepton at Babar applies and limits $\varepsilon$ at the level of $10^{-3}$ for  $20\,{\rm MeV}<m_{Z'}<120\,{\rm MeV}$.
There are other bounds from beam dump experiments (not shown) \cite{beamdump,batell2} that limit $\varepsilon$ at the level of $10^{-3}$ in our parameter space below $m_{Z'}=0.1\,{\rm GeV}$.
The planned SPS target experiment at CERN \cite{SPS} could improve the limits from beam dump experiments further.

\subsection{Direct detection}

\begin{figure}
  \begin{center}
 \includegraphics[height=0.45\textwidth]{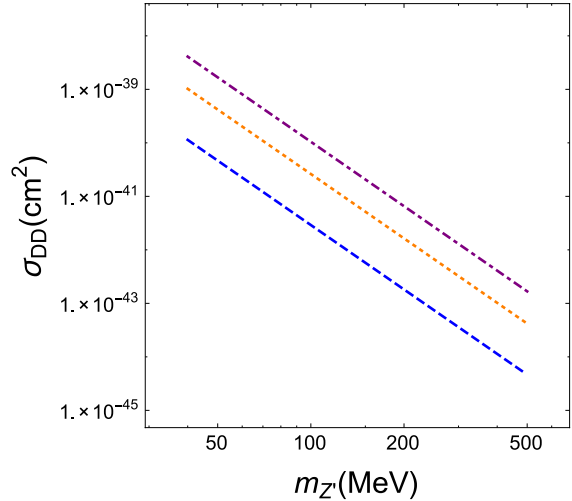}
  \end{center}
  \caption{Direct detection cross section between dark matter and electron as a function of $Z'$ mass. We have chosen $\varepsilon=10^{-4}$, and $g_D=0.1, 0.3, 0.6$ for blue dashed, orange dotted and purple dot-dashed lines, respectively. The cross section depends little on $m_\chi$ in the parameter space satisfying the SIMP relic density.
  }
  \label{dd}
\end{figure}

In the case of $m_e, m_\chi, m_{Z'}\gg p\simeq m_\chi v_{\rm DM}$, which is true of our SIMP dark matter \footnote{When the momentum transfer of order $m_\chi v_{\rm DM}$ is greater than $Z'$ mass, then the DM scattering cross section becomes enhanced due to a Coulomb singularity \cite{zurek}. }, we can take the momenta, $p_1=(m_\chi,p,0), p_2=(m_e,-p,0)$ and $k_1=(m_\chi,p\cos\theta,p\sin\theta)$, $k_2=(m_e, -p\cos\theta,-p\sin\theta)$ in the calculation of the matrix element squared in eq.~(\ref{melements}). 
Then, the DM-electron scattering cross section with $Z'$-portal interaction  is, at present, 
\bea
\sigma_{\rm DD}&=&  \frac{1}{64\pi^2 s} \int d\Omega \,\frac{|{\vec k}_1|}{|{\vec p}_1|}\,|{\cal M}_{\rm scatt}|^2  \nonumber \\
&=& \frac{\varepsilon^2 e^2 g^2_D \mu^2}{\pi m^4_{Z'}}
\eea
where $\mu\equiv m_e m_\chi/(m_e+m_\chi)$ is the reduced mass of the DM-electron system. 
For $m_\chi\gg m_e$, the DM-electron scattering cross section depends little on the DM mass, but is sensitive to the gauge coupling and mass of dark photon.  
The electron recoil energy can be used to search for the signal of a sub-GeV light dark matter \cite{xenon10,zurek}.

In Fig.~\ref{dd}, we depict the DM-electron scattering cross section as a function of the $Z'$ mass by varying the dark photon gauge coupling $g_D$ between $0.1, 0.3, 0.6$ but with fixed gauge kinetic mixing $\varepsilon=10^{-4}$. The result is insensitive to the SIMP mass,  which ranges between $30\,{\rm MeV}$ and $150\,{\rm MeV}$ as shown from the left figure of Fig.~\ref{self}. 
In Fig.~\ref{zprime}, we also present the DM-electron scattering cross section on the parameter space of $\varepsilon$ vs $m_{Z'}$.  The direct search for dark matter with electron recoil in XENON10 \cite{xenon10} can give stronger bounds than beam dump experiments, at the level of $\varepsilon\sim 10^{-1}(10^{-3})$ for $m_{Z'}\sim 0.06\,{\rm GeV}$ with $g_D=0.3(0.6)$. 
XENON10 \cite{xenon10} constrains the DM-electron scattering cross section to be below $2\times 10^{-36}\,{\rm cm}^2$ at  best at about $30\,{\rm MeV}$ so the SIMP parameter space in our model that is compatible with direct $Z'$ searches satisfies the current XENON10 limits. But, the case with a light $Z'$ gauge boson could be also probed by future superconducting detectors \cite{zurek}.

\subsection{Higgs portal couplings and Higgs signals}

We discuss the impact of Higgs portal couplings on the DM scattering and annihilation processes. As shown in the second Feynman diagrams in Fig.~7, the SM Higgs and the dark Higgs contribute to the DM scattering with the SM charged leptons. 
The corresponding annihilation cross section is given by
\be
\langle\sigma v_{\rm rel}\rangle_{\chi f\rightarrow \chi f}=\frac{1}{16\pi m^2_\chi}\Big(\frac{m_f}{v}\Big)^2\left|\frac{y_{h_1\chi^*\chi}}{m^2_{h_1}}+\frac{y_{h_2\chi^*\chi}}{m^2_{h_2}} \right|^2(3m_\chi T_F+m^2_f)\label{hportal1}
\ee
where $T_F= m_\chi/20$ is the freeze-out temperature and $y_{h_1\chi^*\chi}\equiv \sin\theta (\lambda_{\phi\chi}v'\cos\theta-\lambda_{\chi H}v\sin\theta )$
and $y_{h_2\chi^*\chi}\equiv- \cos\theta  (\lambda_{\phi\chi}v'\sin\theta+\lambda_{\chi H}v\cos\theta )$.

On the other hand, the annihilation cross section for $\chi\chi^*\rightarrow f{\bar f}$ is
\be
\langle\sigma v_{\rm rel}\rangle_{\chi\chi^*\rightarrow f {\bar f}}= \frac{1}{4\pi}\Big(\frac{m_f}{v}\Big)^2\Big(1-\frac{m^2_f}{m^2_\chi}\Big)^{3/2}\left|\frac{y_{h_1\chi^*\chi}}{4m^2_\chi-m^2_{h_1}}+\frac{y_{h_2\chi^*\chi}}{4m^2_\chi - m^2_{h_2}} \right|^2. \label{hportal2}
\ee

We digress to the bounds from Higgs signals, which constrain the Higgs mixing angle and the DM coupling to the SM Higgs, $\lambda_{\chi H}$. The bound on the branching fraction of the Higgs invisible decay is ${\rm BR}(h_2\rightarrow \chi\chi^*)<0.29$ at $95\%$ CL from the VBF Higgs production at ATLAS \cite{atlasVBF}. The $ZH$ production at ATLAS \cite{atlasZH} and the VBF$+ZH$ production CMS \cite{cmsHinv} leads to less strong bounds, ${\rm BR}(h_2\rightarrow \chi\chi^*)<0.75$ and ${\rm BR}(h_2\rightarrow \chi\chi^*)<0.58$, respectively. On the other hand, the Higgs signal strength is bounded to $\mu>0.81$ at $95\%$ CL from ATLAS/CMS data combined \cite{falkowski}.   

The partial decay rates for additional Higgs decay modes in our model are
\bea
\Gamma(h_2\rightarrow \chi\chi^*)&=&\frac{y^2_{h_2\chi^*\chi}}{16\pi m_{h_2}}\sqrt{1-\frac{4m^2_\chi}{m^2_{h_2}}}, \\
\Gamma(h_2\rightarrow h_1 h_1)&\simeq& \frac{\lambda^2_{\phi H} v^2}{32\pi m_{h_2}}\sqrt{1-\frac{4m^2_{h_1}}{m^2_{h_2}}}.
\eea
Then, from the bound on the Higgs invisible decay, we obtain $|y_{h_2 \chi^*\chi}|/v\lesssim 0.011$. 
On the other hand, scaling the total and individual Higgs decay rates as well as the Higgs production cross section approximately by $\cos^2\theta$, the Higgs signals bounds the Higgs mixing angle as $\sin\theta<0.44$. In our case, the Higgs mixing angle is approximated to 
$\sin\theta\simeq \lambda_{\phi H} vv'/m^2_{h_2}\lesssim 0.016$ for $m_{h_2}\gg m_{h_1}$, $v'\sim 1\,{\rm GeV}$ and $\lambda_{\phi H}\lesssim 1$. Thus, it is consistent with the Higgs visible decays.  
Given the bounds, $|y_{h_2 \chi^*\chi}|/v\lesssim 0.011$ and $\sin\theta\lesssim 0.016$, we find that the dark Higgs contributes dominantly to the DM scattering and annihilation, as compared to the SM Higgs.
As for $Z'$ portal, we can parametrize the 2-to-2 cross sections with dark Higgs exchanges by $\langle\sigma v\rangle_{\rm ann}=\delta^2_1/m^2_\chi$ and $\langle\sigma v\rangle_{\rm scatt}=\delta^2_2/m^2_\chi$, in eqs.~(\ref{hportal2}) and (\ref{hportal1}), respectively. 
Therefore, taking the charged lepton to be electron or positron \footnote{When DM is heavier than muon, it can annihilate into a muon pair too. But, the DM scattering rate with muon is suppressed by Boltzmann factor at freeze-out \cite{simp2}, because the typical freeze-out temperature is small,  $T_F\sim m_\chi/20\lesssim 20\,{\rm MeV}$. } and choosing $\lambda_{\phi\chi}\sim 1$, we get $\delta_1\sim \delta_2\sim 10^{-10}$.  Therefore, unless $\lambda_{\phi\chi}\gtrsim 10$, the Higgs portal does not satisfy the SIMP condition.

\subsection{Indirect detection}

The SIMP dark matter can be searched for by indirect detection experiments with diffuse gamma-rays. 
Although the Fermi-LAT is sensitive to gamma-rays of energies from $20\,{\rm MeV}$ to $>300\,{\rm GeV}$, its current limits, for instance, from dwarf spheroidal galaxies (dSphs) of the Milky Way \cite{dwarfs}, do not extend down to  $20\,{\rm MeV}$. On the other hand, the AMS positron fraction in primary cosmic rays is within the energy range from 0.5 to 500 GeV \cite{ams02}. The current best upper limits on $e^+e^-(\mu^+\mu^-)$ annihilation cross section is $5\times 10^{-27}(10^{-26}){\rm cm^3/s}$ at $m_\chi=5\,{\rm GeV}$ from Fermi-LAT dSphs \cite{dwarfs} and $10^{-28}(10^{-27}){\rm cm^3/s}$ at similar DM masses from AMS-02 positron fraction \cite{ams02-limit}.  More importantly, there are diffuse X-ray and gamma-ray observations near the center of the galaxy such as INTEGRAL \cite{integral}, COMPTEL \cite{comptel} and EGRET \cite{egret} and Fermi-LAT \cite{fermi-diffuse}.
In particular, COMPTEL leads to the bounds stronger than the thermal annihilation cross section around $3\times10^{-26}\,{\rm cm^3/s}$, for DM masses below $\sim 100\,{\rm MeV}$ \cite{essig}.

SIMP can annihilate into $e^+e^-$ or $\mu^+\mu^-$ depending on its mass, through $Z'$ or Higgs portal with the annihilation cross section,
\be
\langle \sigma v_{\rm rel}\rangle_{l^+l^-}=\Big(1.2\times 10^{-27}{\rm cm^3/s}\Big) 
\left(\frac{\delta_1}{10^{-6}}\right)^2 \left(\frac{100\,{\rm MeV}}{m_\chi}\right)^2.
\ee
In the case of $Z'$ portal, the annihilation cross section is p-wave suppressed as $\delta_1(T)=\delta_{1,F}\sqrt{T/T_F}$ where $\delta_{1,F}=\delta_1(T_F)\lesssim 10^{-6}$ from the  SIMP condition, we get  $\delta_1(T)=(0.01-0.1)\delta_{1,F}\lesssim 10^{-8}-10^{-7}$ at present.   
On the other hand, the annihilation cross section with Higgs-portal is given by $\delta_1=10^{-10}(m_l /m_e)$, which could be as large as $10^{-8}$ for $m_l=m_\mu$. Therefore, in either $Z'$ or Higgs portal, there is no current bound from indirect detection.

\section{Conclusions}

We have shown that models with gauged $Z_3$ symmetry provide a consistent SIMP scalar dark matter with sub-GeV masses that can be in kinetic equilibrium with the SM fermions via $Z'$-portal interactions. 
The resulting large self-scattering cross section of the SIMP dark matter could explain the DM halo separation observed in Abell 3827 cluster, although there is a tension with the previous bounds coming from Bullet cluster and halo shapes. 
Inclusion of $Z'$ gauge boson and dark Higgs changes DM annihilation and scattering processes and opens up a new parameter space of the SIMP self-couplings and masses. 
We found that SIMP conditions are complementary to indirect and collider $Z'$ searches, DM direct/indirect detection and Higgs signals. The SIMP parameter space could be explored further by future collider and indirect/direct detection experiments.

\section*{Acknowledgments}

We would like to thank Jong-Chul Park, Seong Chan Park and Min-Seok Seo for discussion. 
The work of HML is supported in part by Basic Science Research Program through the National Research Foundation of Korea (NRF) funded by the Ministry of Education, Science and Technology (2013R1A1A2007919). The work of SMC is supported by the Chung-Ang University Graduate Research Scholarship in 2015.

\def\theequation{A.\arabic{equation}}

\setcounter{equation}{0}

\vskip0.8cm
\noindent
{\Large \bf Appendix A: Higgs and $Z'$ portal interactions to dark matter} 
\vskip0.4cm
\noindent

After the breakdown of electroweak symmetry and dark $U(1)_V$, we can take the SM Higgs and dark Higgs in unitary gauge,
\be
H=\frac{1}{\sqrt{2}} \left( \begin{array}{c} 0 \\ v+h \end{array} \right), \quad \phi=\frac{1}{\sqrt{2}} (v'+h').
\ee
On the other hand, the VEV of scalar dark matter  $\chi$ is assumed  to vanish. 
Then, the mass matrix for $h', h$ is given by
\be
M = \left( \begin{array}{cc} 2\lambda_\phi v^{\prime 2} & \lambda_{\phi H} v' v \\  \lambda_{\phi H} v'v & 2\lambda_H v^2 \end{array} \right).
\ee
where the VEVs are determined by the following conditions,
\be
v^2=\frac{4\lambda_\phi m^2_H - 2\lambda_{\phi H} m^2_\phi}{4\lambda_H\lambda_\phi -\lambda^2_{\phi H}}, \quad v^{\prime 2}=\frac{4\lambda_H m^2_\phi - 2\lambda_{\phi H} m^2_H} {4\lambda_H\lambda_\phi -\lambda^2_{\phi H}}.
\ee
Then, the conditions for a local minimum are
\bea
&&4\lambda_\phi \lambda_H > \lambda^2_{\phi H}, \quad \lambda_\phi >0, \quad \lambda_H >0,  \\
 && 2\lambda_\phi m^2_H - \lambda_{\phi H} m^2_\phi >0, \\
 && 2\lambda_H m^2_\phi - \lambda_{\phi H} m^2_H >0.
\eea
The mass matrix can be diagonalized by the rotation by
\be
\left( \begin{array}{c} h_1 \\ h_2 \end{array} \right)=  \left( \begin{array}{cc} \cos\theta & -\sin\theta \\  \sin\theta & \cos\theta \end{array} \right) \left( \begin{array}{c} h' \\ h \end{array} \right)
\ee
where $h_1,h_2$ are mass eigenstates. 
The mass eigenvalues of Higgs-like states are
\be
m^2_{h_1,h_2}=\lambda_\phi v^{\prime 2}+\lambda_H v^2 \mp \sqrt{(\lambda_\phi v^{\prime 2}-\lambda_H v^2)^2+ \lambda^2_{\phi H}v^{\prime 2} v^2 }
\ee
and the mixing angle is 
\be
\tan 2\theta = \frac{\lambda_{\phi H} v' v}{\lambda_H v^2-\lambda_\phi v^{' 2}}. 
\ee

Consequently, the self-interactions and Higgs-portal interactions of dark matter are 
\bea
-{\cal L}_{\rm scalar}&=&( \lambda_{\phi \chi} v' \cos\theta -\lambda_{\chi H}v\sin\theta ) h_1 |\chi|^2 +  (\lambda_{\phi \chi} v' \sin\theta + \lambda_{\chi H}v\cos\theta ) h_2 |\chi|^2  \nonumber \\
&&+\frac{1}{2}(\lambda_{\phi \chi}\cos^2\theta+\lambda_{\chi H} \sin^2\theta) h^2_1 |\chi|^2 +\frac{1}{2}(\lambda_{\phi \chi}\sin^2\theta+\lambda_{\chi H} \cos^2\theta) h^2_2 |\chi|^2 \nonumber \\
&&+(\lambda_{\phi\chi}-\lambda_{\chi H}) \sin\theta\cos\theta \, h_1 h_2 |\chi|^2 +\bigg(\frac{1}{3!} \kappa\cos\theta \, h_1 \chi^3++\frac{1}{3!} \kappa\sin\theta \, h_2 \chi^3+{\rm h.c.}\bigg)
 \nonumber \\
 &&+\lambda_\chi |\chi|^4 +\bigg( \frac{1}{3!}\kappa v' \chi^3 +{\rm h.c.}\bigg). 
\eea

On the other hand, the gauge kinetic and mass terms can be diagonalized by
\be
\left(\begin{array}{c} B_\mu \\ W^3_\mu \\ V_\mu \end{array}\right)=\left( \begin{array}{ccc} c_W & -s_W c_\zeta +t_\xi s_\zeta  & -s_W s_\zeta-t_\xi c_\zeta  \\   s_W & c_W c_\zeta & c_W s_\zeta   \\  0 & -s_\zeta/c_\xi & c_\zeta/ c_\xi   \end{array} \right)  \left(\begin{array}{c} A_\mu \\ Z_{1\mu} \\ Z_{2\mu} \end{array}\right)
\ee
where $(B_\mu, W^3_\mu,V_\mu)$ are hypercharge, neutral-weak and dark gauge fields, and $(A_\mu, Z_{1\mu},Z_{2\mu})$ are mass eigenstates, and $s_w\equiv \sin\theta_W, c_W\equiv \cos\theta_W$, etc.  The mass eigenvalues of $Z$-boson and dark photon are
\bea
m^2_{1,2}= \frac{1}{2}\left[m^2_Z (1+s^2_W t^2_\xi)+m^2_V/c^2_\xi\pm \sqrt{(m^2_Z(1+s^2_W t^2_\xi)+m^2_V/c^2_\xi)^2- 4m^2_Z m^2_V /c^2_\xi} \,\right]
\eea
where $m^2_Z=\frac{1}{4}(g^2+g^{\prime 2}) v^2$ and $m^2_V=9g^2_D v^{\prime 2}$, and the mixing angle between $Z$-boson and dark photon is given by
\be
\tan 2\zeta =\frac{m^2_Z s_W \sin 2\xi}{m^2_V-m^2_Z(c^2_\xi-s^2_W s^2_\xi )}.
\ee
As a consequence of the basis rotation, the current interactions can be written as
\bea
{\cal L}&=& e A_\mu J^\mu_{\rm EM} \nonumber \\
&&+ Z_{1\mu} \bigg[ e \varepsilon J^\mu_{\rm EM}+\frac{e}{2s_Wc_W} (c_\zeta-t_W \varepsilon/t_\zeta  )J^\mu_Z -g_D \frac{s_\zeta}{c_\xi} J^\mu_D \bigg] \\
&&+Z_{2\mu}  \bigg[- e \varepsilon J^\mu_{\rm EM}+\frac{e}{2s_Wc_W} (s_\zeta+t_W\varepsilon)J^\mu_Z +g_D \frac{c_\zeta}{c_\xi} J^\mu_D \bigg]
\eea
 where $\varepsilon\equiv c_W t_\xi c_\zeta$, and $J^\mu_{\rm EM}$, $J^\mu_Z$and $J^\mu_D$ are electromagnetic, neutral and dark currents, respectively.    
 For $m_V\ll m_Z$, we get $\zeta\simeq -s_W \xi$, so the neutral current interaction of dark photon is negligible due to $s_\zeta+t_W\varepsilon \simeq \zeta+ s_W\xi\simeq 0$.

\def\theequation{B.\arabic{equation}}

\setcounter{equation}{0}

\vskip0.8cm
\noindent
{\Large \bf Appendix B: General formulas for cross sections} 
\vskip0.4cm
\noindent

The cross sections for 3-to-2 self-scattering, DM-DM annihilation, DM-SM kinetic scattering and 2-to-2 self-scattering, in order, can be written in terms of the squared amplitudes as follows,
\bea
(\sigma v^2_{\rm rel})_{3\rightarrow 2}&=& \frac{\sqrt{5}}{384 \pi m^3_\chi} |{\cal M}_{3\rightarrow 2}|^2, \\
(\sigma v_{\rm rel})_{\rm ann}&=& \frac{1}{32\pi m^2_\chi} |{\cal M}_{\rm ann}|^2, \\
(\sigma v_{\rm rel})_{\rm scatt}&=& \frac{1}{16\pi m^2_\chi} |{\cal M}_{\rm scatt}|^2, \\
\sigma_{\rm self}&=& \frac{1}{64\pi m^2_\chi }|{\cal M}_{\rm self}|^2
\eea
where a $1/2$ factor is taken into account in the squared matrix elements for identical final states and we have ignored the SM particle masses.

\end{document}